\RequirePackage{fix-cm}

\documentclass[smallextended]{svjour3}    
\usepackage{lineno}
\usepackage[xcolor]{changebar}

\usepackage{lipsum}% http://ctan.org/pkg/lipsum

\usepackage{enumitem}
\usepackage[T1]{fontenc}

\usepackage{tikz}
\usepackage{url}
\usepackage{authblk}
\usepackage{mathtools, cuted}
\usepackage{lipani_math}
\usepackage{booktabs} % For formal tables
\usepackage{svg}
\usepackage{comment}
\usepackage{xcolor}
\usepackage{multirow}
\usepackage{multicol}
\usepackage{enumitem}
\usepackage{fontawesome}
\usepackage{adjustbox}
\usepackage{graphicx,mathtools}
\usepackage{supertabular}
\usepackage[labelfont=bf]{caption}
\usepackage{amsmath}
\usepackage{amsthm}
\usepackage{bm}
\usepackage[misc,geometry]{ifsym}

\usepackage{academicons}
\newcommand{\orcid}[1]{\href{https://orcid.org/#1}{\textcolor[HTML]{A6CE39}{\aiOrcid}}}

\usepackage{breqn}
\usepackage[pdftex,breaklinks]{hyperref}
\usepackage[round]{natbib}

\newcommand{\sizecirle}{0.8ex}
\newcommand{\rc}{\tikz\fill[red] (0,0) circle (\sizecirle);}
\newcommand{\bc}{\tikz\fill[blue] (0,0) circle (\sizecirle);}
\newcommand{\tc}{%
\begin{tikzpicture}
\fill[red] (0,0) circle (\sizecirle); 
\fill[blue] (0,0) -- (90:-\sizecirle) arc (-90:90:\sizecirle) -- cycle;
\fill[white] (0,0) circle (0.5ex); 
\end{tikzpicture}}
\newcommand{\wc}{%
\begin{tikzpicture}
\fill[red] (0,0) circle (\sizecirle); 
\fill[blue] (0,0) -- (90:-\sizecirle) arc (-90:90:\sizecirle) -- cycle;
\fill[white] (0,0) circle (0.5ex); 
\end{tikzpicture}}

\newcommand{\lnr}{\text{\faClose}}

\newcommand{\lps}{\text{\faThumbsOUp}}
\newcommand{\lns}{\text{\faHandRockO}}
\newcommand{\las}{\text{\raisebox{-2pt}{\faThumbsODown}}}

\newcommand\correspondingauthor{\thanks{\Letter\hspace{0.3cm}Gizem Gezici \\ gizemgezici@sabanciuniv.edu}}

\DeclareRobustCommand{\llp}{\text{\bc}}
\DeclareRobustCommand{\lcp}{\text{\rc}}
\DeclareRobustCommand{\lbp}{\text{\tc}}

\begin{document}

\title{Evaluation Metrics for Measuring Bias in Search Engine Results}
\titlerunning{Evaluation Metrics for Measuring Bias in Search Engine Results}
%\title{Of Bias and Search Engines: Evaluating Search Engine Results Through Rank and Relevance Based Measures}
%\titlerunning{Evaluating Search Engine Results Through Rank and Relevance Based Measures}
%Biased or Not? The Story of Two Unaccountable Search Engines]{Biased or Not? \\The Story of Two Unaccountable Search Engines}% or Not? \\The Story of Two Unaccountable Search Engines\vspace{1em}}
%\titlenote{Produces the permission block, and copyright information}
%\subtitle{Extended Abstract}
%\subtitlenote{The full version of the author's guide is available as \texttt{acmart.pdf} document}

%
%
%\authorrunning{G. Gezici et al.}
% First names are abbreviated in the running head.
% If there are more than two authors, 'et al.' is used.
%

%\author[1]{Author A}
%\author[2]{Author D}
%\author[1,2]{Author E}

%\author{Gizem Gezici\thanks{XYZ} \and DEF\thanks{UVW} \and GHI\footnotemark[1]}
%\footnote[2]{University College London, Department of Computer Science, London, UK}

\author[1]{Gizem Gezici\correspondingauthor{}}
\author[2]{Aldo Lipani}
\author[1]{Yucel Saygin}
\author[2]{Emine Yilmaz}
\affil[1]{Sabanci University, Department of Computer Science and Engineering, Istanbul, Turkey}
\affil[2]{University College London, Department of Computer Science, London, UK}

%\myauthor{Author One}

\authorrunning{G.~Gezici et al.}

\date{Received: 4 February 2020 / Accepted: 7 December 2020}
\maketitle

\begin{abstract}
Search engines decide what we see for a given search query. 
Since many people are exposed to information through search engines, it is fair to expect that search engines are neutral. However, search engine results do not necessarily cover all the viewpoints of a search query topic, and they can be biased towards a specific view since search engine results are returned based on relevance, which is calculated using many features and sophisticated algorithms where search neutrality is not necessarily the focal point. Therefore, it is important to evaluate the search engine results with respect to bias. In this work we propose novel web search bias evaluation measures which take into account the rank and relevance. We also propose a framework to evaluate web search bias using the proposed measures and test our framework on two popular search engines based on 57 controversial query topics such as abortion, medical marijuana, and gay marriage. We measure the \emph{stance bias} (in support or against), as well as the %\emph{political bias} (conservative, neutral or liberal). 
\emph{ideological bias} (conservative or liberal). We observe that the stance does not necessarily correlate with the ideological leaning, e.g. a positive stance on abortion indicates a liberal leaning but a positive stance on Cuba embargo indicates a conservative leaning. Our experiments show that neither of the search engines suffers from stance bias. However, both search engines suffer from ideological bias, both favouring one ideological leaning to the other, which is more significant from the perspective of polarisation in our society. 

\end{abstract}


%Article Title: Measuring Search Engines' Political Bias in Controversial Topics via Stance Bias
%Journal Name: Information Retrieval Journal
%Authors: Gizem Gezici, Aldo Lipani, Yucel Saygin, and Emine Yilmaz
%Corresponding Author Information: Gizem Gezici, Sabanci University, gizemgezici@sabanciuniv.edu

\section{Introduction}
\label{sec:introduction}
%Article Title: Measuring Search Engines' Political Bias in Controversial Topics via Stance Bias
%Journal Name: Information Retrieval Journal
%Authors: Gizem Gezici, Aldo Lipani, Yucel Saygin, and Emine Yilmaz
%Corresponding Author Information: Gizem Gezici, Sabanci University, gizemgezici@sabanciuniv.edu

Search engines have become an indispensable part of our lives. As reported by SmartSights~\citeyearpar{SmartSights}, 46.8\% of the world population accessed the internet in 2017 and by 2021, the number is expected to reach 53.7\%. According to InternetLiveStats~\citeyearpar{InternetLiveStats}, currently on average 3.5 billion Google searches are done per day. These statistics indicate that search engines replaced traditional broadcast media and have become a \emph{major} source of information ``gatekeepers to the Web'' for many people%, especially in the last decade
~\citep{diaz2008through}. As information seekers search the Web more, they are also more influenced by Search Engine Result Pages (SERPs), pertaining to a wide range of areas (e.g., work, entertainment, religion, and politics). For instance, in the course of elections, it is known that people issue repeated queries on the Web about political candidates and events such as ``democratic debate'', ``Donald Trump'', ``climate change''~\citep{kulshrestha2018search}. SERPs returned in response to these queries may influence the voting decisions as claimed by~\citet{epstein2015search}, who report that \textit{manipulated} search rankings can change the voting preferences of undecided individuals at least by 20\%.

Although search engines are widely used for seeking information, the majority of online users tend to believe that they provide \emph{neutral} results, i.e. serving only as facilitators in accessing information on the Web~\citep{goldman2008search}. However, there are counter examples to that belief as well. A recent dispute between the U.S.~President Donald Trump and Google is such an example, where Mr.~Trump accused Google of displaying only negative news about him when his name is searched to which Google responded by saying: ``When users type queries into the Google Search bar, our goal is to make sure they receive the most relevant answers in a matter of seconds'' and ``Search is not used to set a political agenda and we don't bias our results toward any political ideology''~\citep{ReutersTrump}. %\footnote{https://www.reuters.com/article/us-usa-trump-tech-alphabet/google-responds-to-trump-says-no-political-motive-in-search-results-idUSKCN1LD1QP}.
In this work, we hope to shed some light on that debate, by not specifically concentrating on queries regarding Donald Trump but by conducting an in depth analysis of search answers to a broad set of controversial topics based on concrete evaluation measures.

\emph{Bias} is defined with respect to balance in representativeness of Web documents retrieved from a database for a given query~\citep{mowshowitz2002assessing}. When a user issues a query to a search engine, %the SERPs 
documents from different sources are gathered, ranked, and displayed to the user. Assume that a user searches for \emph{2016 presidential election} and the top-n ranked results are displayed. In such a search scenario, the retrieved results may favor some political perspectives over others and thereby fail to provide impartial knowledge for the given query as claimed by Mr.~Trump, though without any scientific support. 
Hence, the potential \emph{undue emphasis} of specific perspectives (or viewpoints) in the retrieved results lead to bias~\citep{kulshrestha2018search}. With respect 
to the definition of bias and the presented scenario, if there is an unbalanced representation, i.e. skewed or slanted distribution, of the viewpoints in a SERP, i.e. not only in political searches, towards the query's topic, then we consider this SERP as \emph{biased} for the given search query. %In this work, we initially analyse the bias in terms of stances where \emph{stance} is defined as the document is in favour of the given target or against it. We further analyse the ideological bias by interpreting the document stances in conservative-liberal ideology space and transforming these stances into ideological leanings, i.e. conservative or liberal. and ideologies in search results from various domains. , whereas \emph{ideology} stands for a collection of ideas or beliefs shared by a group of people.
%of the perspectives towards the query's topic, implied by a skewed or slanted set of SERPs, then we consider these SERPs as \emph{biased} for the given search query.
%

Bias is especially important if the query topic is \emph{controversial} having opposing views, in which case it becomes more critical that search engines are supposed to return results with a \emph{balanced} representation 
%an \emph{even} split, i.e. unbiased results, 
of different perspectives which implies that they do not favour one specific perspective %(or candidate) 
over another. Otherwise, %in the case of \emph{controversial} topics 
this may dramatically affect public as in the case of elections leading to polarisation in society for \emph{controversial} issues. On the other hand, returning an unbalanced representation of distinct viewpoints is not sufficient to claim that the search engine's ranking algorithm is biased. One reason for a skewed SERP could be due to the corpus itself, i.e. if documents indexed and returned for a given topic come from a slanted distribution, meaning that the ranking algorithm returns a biased result set due to a biased corpus. 
To differentiate the algorithmic vs corpus bias, one needs to investigate the source of bias in addition to the skewed list analysis of the top-n search results. However, the existence of bias, regardless of being corpus or algorithmic bias, would still conflict with the expectation that an IR system should be fair, accountable, and transparent~\citep{culpepper2018research}. %Also, 
Furthermore, it was reported that people are more susceptible to bias when they are unaware of it~\citep{bargh2001automated}, and \citet{epstein2017suppressing} showed that alerting users about bias can be effective in suppressing search engine manipulation effect (SEME). %Thus, at least the bias information should be used by search engines to inform their users about the bias and decrease the possible SEME by making themselves more accountable, thereby alleviating the negative effects of bias and serving only as facilitators as they generally claim of.
Thus,  search engines should at least inform their users about the bias and decrease the possible SEME by making themselves more
accountable, thereby alleviating the negative effects of bias and serving only as facilitators as they generally claim to be. In this work, we aim to serve that purpose by proposing a search bias evaluation framework taking into account the rank and relevance
~\footnote{We are referring to the notion of relevance defined in the literature as system relevance, or topical relevance which is the relevance predicted by the system.}
of the SERPs. Our contributions in this work can be summarised as follows:

\begin{enumerate}
  \item We propose a \emph{new generalisable search bias evaluation framework} to measure bias in SERPs by quantifying two different types of bias on content which are stance bias and ideological bias.
  \item We present \emph{three novel fairness-aware measures of bias} that do not suffer from the limitations of the previously presented bias measures, based on common Information Retrieval (IR) \emph{utility-based} evaluation measures: Precision at cut-off (P$@n$), Rank Biased Precision (RBP), and Discounted Cumulative Gain at cut-off (DCG$@n$) which are explained in Section~\ref{sec:measuresbias} in detail.% While the first measure quantifies bias considering only a weak ranking criterion, i.e.~the first $n$ documents as in SERPs, the other two measures incorporate stronger ranking bias.
  \item We apply the proposed framework to \emph{measure the stance and ideological bias} not only in political searches but searches related to a wide range of controversial topics; including but not limited to education, health, entertainment, religion and politics on Google and Bing \emph{news} search results.
  \item We also utilise our framework to \emph{compare the relative bias} for queries from various controversial issues on two popular search engines: Google and Bing news search.
\end{enumerate}

We would like to note that we distinguish the stance and ideological leaning in SERPs. The stance in a SERP for a query topic could be in favor or against the topic, whereas the ideological leaning in a SERP stands for the specific ideological group as conservatives or liberals that supports the corresponding topic. Hence, the stance in a SERP does not directly imply the ideological leaning. For example, given two controversial queries, "abortion" and "Cuba embargo", a SERP could have a positive stance for the topic of abortion, indicating a liberal leaning, while a positive stance for the topic of Cuba embargo indicates a conservative leaning. Therefore looking at the stance of the SERPs for controversial issues is not enough and could even be misleading in determining the ideological bias. We demonstrate how the proposed framework can be used to quantify bias in the SERPs of search engines (in this case Bing and Google) in response to queries related to \emph{controversial} topics. Our analysis is mainly two-fold where we first evaluate stance bias in SERPs, and then use this evaluation as a proxy to quantify ideological bias asserted in the SERPs of the search engines. 

In this work, via the proposed framework, we aim to answer the following research questions: 
\begin{description}
    \item[RQ1:] On a pro-against stance space, do search engines return \emph{biased} SERPs towards controversial topics? 
    \item[RQ2:] Do search engines show \emph{significantly different} magnitude of stance bias from each other towards controversial topics?
    \item[RQ3:] On a conservative-liberal ideology space, do search engines return \emph{biased} SERPs and if so; are these biases \textit{significantly different} from each other towards controversial topics? %Do we reach the same conclusions when transforming the pro-to-against scale to conservative-to-liberal? %are search engines \textit{significantly different} from each other towards controversial topics?
\end{description}
We address these research questions for controversial topics representing a broad range of issues in SERPs of Google and Bing through content analysis, i.e. analysing the textual content of the retrieved documents. 
In order to answer RQ1, we measure the degree of deviation of the ranked SERPs from an \emph{ideal} distribution, where different stances are \emph{equally} likely to appear. To detect bias which results from the unbalanced representation of distinct perspectives, we label the documents' stances with crowd-sourcing and use these labels for stance bias evaluation.
In this paper we focus on a particular kind of bias, \emph{statistical parity} or more generally known as \emph{equality of outcome}, i.e. given a population divided into groups, the groups in the output of the system should be equally represented. This is in contrast with the other popular measure generally known as \emph{equality of opportunity}, i.e. given a population divided into groups, the groups in the output should be represented based on their proportion in the population namely, base rates. For choosing the \emph{equality of outcome}, we have mainly two reasons. First, in the context of the controversial topics, not all of the corresponding debate questions (queries) have certain answers based on scientific facts. Second, the identification of the stance for the full ranking list, i.e. which is a fair representative set of the indexed documents, is too expensive to get annotated through crowd-sourcing. Thus, this choice of \emph{ideal} ranking makes the experiments feasible.
%To detect bias which results from the imbalanced representation of distinct opinions, we label the documents' stances with crowd-sourcing and use these labels for stance bias evaluation.
To address RQ2, we compare the stance bias in the SERPs of the two search engines to see if they show similar level of bias for the corresponding controversial topics. 

%\cbstart
%Example change bar
%This the text which has updates, or changed, or added etc.

%\cbend

For instance, the topic of \emph{abortion} has the query of \emph{Should Abortion Be Legal?} Since mostly liberals support the proposition in this query, liberal leaning is assigned to abortion. The stance labels of the retrieved
documents towards the query are transformed into ideological leanings as follows. If a document has the pro stance which means that it supports the asserted proposition, then its ideological leaning is liberal; if it has the against stance, its leaning is conservative.

%In the next page, the workers are shown a HIT with a topic question (query), i.e. one of the main debates of the corresponding topic, and asks the worker the following:Which political group would answer favourably to this question?. The topics assigned to conservative or liberal perspectives have been decided based on the judgment of five annotators with majority-voting. 
%Additionally, we requested the workers to evaluate the perspective of a given topic based on the current political climate.

In our bias evaluation framework, we concentrate on the top-10 SERPs coming from the~\emph{news} sources to investigate two major search engines (Bing and Google) in terms of bias. We deliberately use~\emph{news} SERPs for our experiments since they often exhibit a specific view towards a topic~\citep{alam2014analyzing}.
%orientation towards a concept
Recent studies~\citep{ZeroLimit, 99Firms} show that on average more than 70\% of all the clicks are in the first page results, thus we only focus on the top-10 results to show the existence of bias. Experiments show that there is no statistically significant difference of \emph{stance bias} in magnitude measured across the two search engines, meaning that they do not favour one specific stance over other. 
However, we should stress that stance bias results need to be taken with a grain of salt as demonstrated through the abortion and Cuba embargo query examples. Polarisation of the society is mostly on ideological leanings, and our second phase of experiments show that there is statistically significant difference of \emph{ideological bias}, where both search engines favour one ideological leaning over other.   

The remainder of the paper is structured as follows. %In Section \ref{sec:notation}, we present the notation used throughout the paper for clarity and 
In Section \ref{sec:related_work} we give the related work and the search bias evaluation framework is proposed in Section \ref{sec:bias_evaluation_measures}. 
In Section \ref{sec:experiments} we detail the experimental setup, and present the results. Then, we discuss the results in Section \ref{sec:discussion}. In Section \ref{sec:limitations} we present the limitations of this work, and we conclude in Section \ref{sec:conclusion}.

%\section{Notation}
%\label{sec:notation}
%\input{notation.tex}

\section{Background \& Related Work}
\label{sec:related_work}
%Article Title: Measuring Search Engines' Political Bias in Controversial Topics via Stance Bias
%Journal Name: Information Retrieval Journal
%Authors: Gizem Gezici, Aldo Lipani, Yucel Saygin, and Emine Yilmaz
%Corresponding Author Information: Gizem Gezici, Sabanci University, gizemgezici@sabanciuniv.edu

In recent years, bias analysis in SERPs of search engines has attracted a lot of interest~\citep{baeza2016data,mowshowitz2002bias,noble2018algorithms,pan2007google,tavani2012search} due to the concerns that search engines may manipulate the search results influencing users. The main reason behind these concerns is that search engines have become the fundamental source of information~\citep{dutton2013cultures}, and surveys from Pew~\citeyearpar{american2014personal} and Reuters~\citeyearpar{newman2018reuters} found that more people obtain their news from search engines than social media. The users reported higher trust on search engines for the accuracy of information~\citep{newman2018reuters, newman2019reuters, PewTrust} and many internet-using US adults even use search engines to fact-check information~\citep{dutton2017search}. 

To figure out how this growing usage of search engines and trust in them might have undesirable effects on public, and what could be the methods to measure those effects, in the following we review the research areas related first to automatic stance detection, then to fair ranking evaluation, and lastly to search bias quantification.

%\subsection{Opinion Mining} %Topic Discovery \& Sentiment Analysis}

\subsection{Opinion Mining and Sentiment Analysis}

A form of Opinion Mining related to our work is Contrastive Opinion Modeling (COM). Proposed by~\citet{fang2012mining}, in COM, given a political text collection, the task is to present the opinions of the distinct perspectives on a given query topic and to quantify their differences with an unsupervised topic model. COM is applied on debate records and headline news. Differently from keyword analysis to differentiate opinions using topic modelling, we compute different IR metrics from the content of the news articles to evaluate and compare the bias in the SERPs of two search engines. 
\citet{aktolga2013sentiment} consider the sentiment towards controversial topics and propose different diversification methods based on the topic sentiment. Their main aim is to diversify the retrieved results of a search engine according to various sentiment biases in blog posts rather than measure bias in the SERPs of~\emph{news} search engines as we do in this work.

\citet{demartini2010dear} exploit automatic and lexicon-based text classification approaches, Support Vector Machines and SentiWordNet respectively to extract sentiment value from the textual content of SERPs in response to controversial topics. %Then, only this sentiment information is used to detect and compare opinions in the retrieved results of three commercial search engines without measuring bias as we present in this study.
Unlike us,~\cite{demartini2010dear} only use this sentiment information to compare opinions in the retrieved results of three commercial search engines without measuring bias.
In this paper, we propose a new bias evaluation framework with robust bias measures to systematically measure bias in SERPs. 
\citet{chelaru2012querysent} focus on queries rather than SERPs and investigate if the opinionated queries are issued to search engines by computing the sentiment of suggested queries for controversial topics. In a follow-up work \citep{chelaru2013querysentfollow}, authors use different classifiers to detect the sentiment expressed in queries and extend the previous experiments with two different use cases. %Since our work analyses the SERPs and focuses on the news domain, we identify the stance of the news articles towards a topic via crowd-sourcing. %In our work, we use opinion mining to detect different viewpoints in SERPs in order to quantify bias.
Instead of queries, our work analyses the SERPs in~\emph{news} domain, therefore we need to identify the stance of the news articles. Automatically obtaining article stances is beyond the scope of this work, thus we use crowd-sourcing.
%Since  instead of queries, Our work analyses the news SERPs instead of queries, therefore to identify the stance of the news articles we use crowd-sourcing 

%Queries may contain inherent polarity as SERPs do. According to~\cite{koutra2015events}, users tend to phrase queries in a way that confirms their own beliefs such as ``Obama born in kenya''. Moreover, selected keywords in a query may imply a perspective as in ``gun control'' vs. ``gun rights''. Therefore, we experimented with and without \emph{polarized} queries in our query set.

\subsection{Evaluating Fairness in Ranking}
\label{subsec:fairness_evaluation}
Fairness evaluation in ranked results has attracted attention in recent years.
\citet{yang2017measuring} propose three bias measures, namely Normalized discounted difference (rND), Normalized discounted Kullback-Leibler divergence (rKL) and Normalized discounted ratio (rRD) that are related to Normalized Discounted Cumulative Gain (NDCG) through the use of logarithmic discounting for regularization which is inspired from NDCG as also stated in the original paper.
%for ranked outputs to quantify fairness 
%in a ranking scheme %Moreover, a data generation procedure is developed to systematically control the degree of unfairness in the output and evaluate the proposed measures. % on synthetically generated and the real datasets.
%In this fairness quantification framework, 
Researchers use these metrics to check if there exists a systematic discrimination against a group of individuals, when there are only two different groups as a protected ($g_1$) and an unprotected group ($g_2$) in a ranking. In other words, researchers quantify the relative representation of $g_1$ (the protected group), whose members share a characteristic such as race or gender that cannot be used for discrimination, in a ranked output.
%These proposed measures have the following shape: %For example, rND is defined as follows:
The definitions of these three proposed measures can be rewritten as follows:
\begin{linenomath*}
\begin{equation}
\label{eq:yang_1}
    f_{g_1}(r) = \frac{1}{Z} \sum_{i=10, 20, \dots}^{|r|}
    \frac{1}{\log_2i}%{\log_2(i+1)}
    \left|d_{g_1}(i,r)\right|,
\end{equation}
\end{linenomath*}
where $f(r)$ is a general definition of an evaluation measure for a given ranked list of documents, i.e. a SERP, whereas $f_{g_1}$ is specifically for the protected group of $g_1$. In this definition, $Z$ is a normalisation constant, $r$ is the ranked list of the retrieved SERP and $|r|$ is the size of this ranked list, i.e. number of documents in the ranked list. 
Note that, $i$ is deliberately incremented by 10, to compute \emph{set-based fairness} at discrete values as top-10, top-20 etc., instead of 1 as usually done in IR for the proposed measures to show the correct behaviour with bigger sample sizes. The purpose of computing the~\emph{set-based fairness} to express that being fair at higher positions of the ranked list is more important, e.g. top-10 vs. top-100.

In the rewritten formula, $d_{g_1}$ defines a distance function between the expected probability to retrieve a document belonging to $g_1$, i.e. in the overall population, and its observed probability at rank $i$ to measure the systematic bias. These probabilities turn out to be equal to P@n:
\begin{linenomath*}
\begin{equation}\label{eq:pc}
    \text{P}_{g_1}@n=\frac{1}{n}\sum_{i=1}^n[j(r_i) = g_1],
\end{equation}
\end{linenomath*}
when computed over $g_1$ at cut-off value $|r|$ and $i$ for the three proposed measures as below. In this formula, $n$ is the number of documents considered in $r$ as a cut-off value, and $r_i$ is defined as the document in $r$ retrieved at rank $i$. Note that, $j(r_i)$ returns the label associated to the document $r_i$ specifying its group as $g_1$ or $g_2$. Based on this, $[j(r_i) = g_1]$ refers to a conditional statement which returns 1 if the document $r_i$ is the member of $g_1$ and 0 otherwise.
%
%where $d(i, r)$ represents a distance between the input distribution we find at rank $i$ and the expected distribution in the collection of documents. 
%
\noindent In the original paper, $d_{g_1}$ is defined for %{\color{red} add the full names of these measures} 
rND, rKL, and rRD as:

\begin{linenomath*}
\begin{align}
    d_{g_1}(i,r) = & \text{P}_{g_1}@i - \text{P}_{g_1}@{|r|} \; & \text{for rND},\\
    d_{g_1}(i,r) = & - \text{P}_{g_1}@i \log \left(\frac{\text{P}_{g_1}@{|r|}}{\text{P}_{g_1}@i}\right) \\ 
    & \qquad - (1-\text{P}_{g_1}@i) \log \left(\frac{1-\text{P}_{g_1}@{|r|}}{1-\text{P}_{g_1}@i}\right) \; & \text{for rKL},\\
    d_{g_1}(i,r) = & \frac{\text{P}_{g_1}@i}{1-\text{P}_{g_1}@i} - \frac{\text{P}_{g_1}@{|r|}}{1-\text{P}_{g_1}@{|r|}} \; & \text{for rRD}.
\end{align}
\end{linenomath*}
\\\\

These measures, although inspired by IR evaluation measures, particularly in the context of content bias in search results suffer from the following limitations:
\begin{enumerate}
\item rND measure focuses on the protected group ($g_1$). If we were to compute $f$ at steps of $1$ with the given equal desired proportion of the two groups as 50:50, then the distance function of rND, denoted as $d_{g_1}$ would always give a value of 0.5 for the first retrieved document, where $i=1$. This will always be the case, no matter which group this document belongs to, e.g. \emph{pro} or \emph{against} in our case. This is caused by $d_{g_1}$ of rND through the use of its absolute value in Eq.~\eqref{eq:yang_1}. In our case, this holds when $i=1, 2, 4$ and $r = 10$ where we measure bias in the top-10 results. This is in fact avoided in the original paper~\citep{yang2017measuring} by computing $f$ at steps of $10$ as top-10, top-20 etc. rather than the steps of $1$ as it is usually done in IR which gives more meaningful results in our evaluation framework.

\item rKL measure cannot differentiate between biases of equal magnitude, but in opposite directions with the given equal desired proportion of the two groups as 50:50, i.e. it cannot differentiate bias towards \emph{conservative}, or \emph{liberal} in our case. Also, in IR settings it is not as easy to interpret the computed values from the KL-divergence (denoted as $d_{g_1}$ for rKL) compared to our measures since our measures are based on the standard utility-based IR measures. %which directly compute the bias values per document in top-10 of the search results for search bias evaluation. 
Furthermore, KL-divergence tends to generate larger distances for small datasets, thus it could compute larger bias values in the case of only $10$ documents, and this situation may become even more problematic if we measure bias for less number of documents, e.g. top-3, top-5 for a more fine-grained analysis. In the original paper, this disadvantage is alleviated by computing the rKL values also at discrete points of steps $10$ instead of $1$.

\item rRD measure does not treat the protected and unprotected groups ($g_1$ and $g_2$) symmetrically as stated in the original paper, which is not applicable to our framework. Our proposed measures treat $g_1$ and $g_2$ equal since we have two protected groups; \emph{pro} and \emph{against} for stance bias, \emph{conservative} and \emph{liberal} for ideological bias to measure bias in search settings. Moreover, rRD is only applicable in special conditions when $g_1$ is the minority group in the underlying population as also declared by the authors, while we do not have such constraints for our measures in the scope of search bias evaluation.

\item These measures focus on differences in the relative representation of $g_1$ between distributions. Therefore, from a general point of view, most probably more samples are necessary for these measures to show the expected behavior and work properly. In the original paper, experiments are fulfilled with three different datasets, one is synthetic which includes 1000 samples and two are real datasets which include 1000 and 7000 samples to evaluate bias with these measures, while we have only 10 samples for query-wise evaluation. This is probably because these measures were mainly devised for the purpose of measuring bias in ranked outputs instead of search engine results; none of these datasets contain search results either. 
\item These measures are difficult to use in practice, since they rely on a normalization term, $Z$ that is computed stochastically, i.e. as the highest possible value of the corresponding bias measure for the given number of documents $n$ and protected group size |$g_1$|. In this paper, we rely on standard statistical tests, since they are easier to interpret, provide confidence intervals, and have been successfully used to investigate inequalities in search systems previously by~\citet{chen2018investigating}.
\item These measures do not consider relevance which is a fundamental aspect when evaluating bias in search engines. For example, as in our case, when searching for a controversial topic, if the first retrieved document is about a news belonging to $g_1$ but its content is not relevant to the searched topic, then these measures would still consider this document as positive for $g_1$. However, this document has absolutely no effect on providing an unbiased representation of the controversial topic to the user. This is because these metrics were devised particularly for evaluating bias in the ranked outputs instead of SERPs.
\end{enumerate}
Although the proposed measures by~\citet{yang2017measuring} are valuable in the context of measuring bias in ranked outputs where the individuals are being ranked and some of these individuals are the members of the protected group ($g_1$), these measures have the aforementioned limitations. These limitations are particularly visible for content bias evaluation where the web documents are being ranked by search engines in a typical IR setting. In this paper we address these limitations by proposing a family of fairness-aware measures with the main purpose of evaluating content bias in SERPs, based on standard utility-based IR evaluation measures.
\\
\indent\citet{zehlike2017fa}, based on \citet{yang2017measuring}'s work, %aim to achieve statistical parity in ranked output through connecting the creation of ranking with assessing fairness metrics. 
propose an algorithm to test the statistical significance of a fair ranking.
\citet{beutel2019fairness} propose a pairwise fairness measure for recommender systems. %Similarly to our work the authors follow the equality of opportunity principle by \cite{hardt2016equality}. 
%Nonetheless, their primary focus is on under-recommended items, whereas we do not focus on a specific group. 
However, the authors, unlike us, measure fairness on personalized recommendations and %but we work in an unpersonalized information retrieval setting and unlike us they also 
do not consider relevance, while we work in an unpersonalized information retrieval setting and we do consider relevance. % labels are known for each item in a ranking.
%
%\citet{kallus2019fairness} studies the cross-ROC curve and the corresponding xAUC metric for auditing disparities in a bipartite ranking setting, which is described as finding a good ranking function to rank positively labeled examples above negative ones. Unlike \cite{kallus2019fairness}, %which focuses on the bipartite ranking setting where the AUC loss mainly evaluates the ranking quality of the entire distribution, our method is concerned with measuring fairness in a search engine by analyzing a fair representation of a political perspective in SERPs.% and our proposed metrics such as DCG-based and top-k metrics evaluate only a portion of the distribution, i.e., returned ranked documents, to detect bias.
\citet{kallus2019fairness} investigate the fairness of predictive risk scores as a bipartite ranking task, where the main goal is to rank positively labelled examples above negative ones. 
%Whilst, in this paper we aim to measure fairness in ranking % as an information retrieval application, i.e., 
%in terms of fair representation in SERPs, instead of the bipartite ranking setting. 
However, their measures of bias based on the area under the ROC curve (AUC) are agnostic from the rank position at which a document has been retrieved. %DCG-based and top-k measures, which emphasize the position of an item in a portion of the search results, the researchers in \cite{kallus2019fairness} present an AUC-variant metric that focuses on ranking quality of the entire search result.
%the synthetic ranking generation procedure of \cite{yang2017measuring} is used to calibrate the proposed algorithm, FA*IR, and optimize the utility of a ranking. The focus of \cite{zehlike2017fa}

%In our work, we do not require the generation of \emph{randomized ranking}, therefore not relying on any other search engine. 

%The work proposed by~\citet{demartini2010dear} can be considered as relevant to ours since supervised and lexicon-based text classification techniques were exploited to detect document and sentence-level sentiment in the textual content where the authors analysed 14 controversial queries for three search engines. However, different from our work, their aim is not to measure bias, but to compare the opinions represented in the SERPs using only average sentiment scores and visualisation.

\subsection{Quantifying Search Engine Biases}
Although the search engine algorithms are not transparent and available to external researchers, algorithm auditing techniques provide an effective means for systematically evaluating the results in a controlled environment~\citep{sandvig2014auditing}.
Prior works leverage LDA-variant unsupervised methods and crowd-sourcing to analyse bias in content, or URL analysis for indexical bias.

%Researchers quantify bias (partisanship) in US news outlets (newspapers and 2 political blogs) for 15 selected queries related to a wide range of controversial issues about which Democrats and Republicans argue. They used a combination of machine-learning and crowdsourcing techniques for content-based analysis. 
%***"Finally, news organizations express their ideological bias not by directly advocating for a preferred political party, but rather by disproportionately criticizing one side, a convention that further moderates overall differences." (Interesting)
%Articles are represented with 1000-dimensional vector and for the labels of them they used AMT for the news and politics classifiers. Quantifying bias için crowdsourcing'i ilk defa bu ekip kullanıyor. Hem topic (LDA den alınan suggested two topics) hem de ideological slant'lerini AMT'de label'latıyorlar bir sample'ın.
%Daha önce audience-based ve content-based bias çalışmaları vardı. Ama audience-based provides only relative not absoulute measures of slant. Bunlar da "content-based" yapıyorlar eğer ikiye ayırırsak yoksa aslında rater-based yapıyorlar content-based altında crowdsourcing kullandıkları için.
%Result: They found that US news outlets are substantially more similar fro lect to right-and less partisan-than generally believed. No strong liberal bias exists in the US media.
%\textcolor{red}{Check here}

%First: Content analyis 
%i.LDA-variant unsupervised methods
\citet{saez2013social} propose unsupervised methods to characterise different types of biases in online news media and in their social media communities by also analysing political perspectives of the news sources.
\citet{yigit2016towards} investigate media bias by analysing the user comments along with the content of the online news articles to identify the latent aspects of two highly polarising topics in the Turkish political arena. \citet{kulshrestha2017quantifying} quantify bias in social media by measuring the bias of the author of a tweet, while in \citet{kulshrestha2018search}, bias in web search is quantified through a URL analysis for Google in political domain without any SERP content analysis. In our work, we consider the Google and Bing SERPs from news sources such as NY-Times, and BBC news in order to quantify bias through content analysis.

%ii.Crowd-sourcing
In addition to the unsupervised approaches, crowd-sourcing is a widely used mechanism to analyse bias in content.
Crowd-sourcing is a common approach for labelling tasks in different research areas such as image \& video annotation \citep{crowd-image,crowd-video}, object detection \citep{crowd-object}, named entity recognition \citep{crowd-ner,crowd-twitter}, sentiment analysis \citep{crowd-sentiment} and relevance evaluation \citep{crowd-relevance, crowd-trec}. \citet{crowd-survey} provide a detailed survey of crowd-sourcing applications. As \citet{crowd-survey} suggest, crowd-sourcing can also be used for gathering opinions from the crowd. \citet{crowd-opinion} use crowd-sourcing to classify Spanish consumer comments and show that non-expert Amazon Mechanical Turk (MTurk) annotations are viable and cost-effective alternative to expert ones. In this work, we use crowd-sourcing for collecting opinions of the public not about consumer products but controversial topics.

%Second: URL analysis for indexical bias
Apart from the content bias, there is another research area, namely indexical bias.
Indexical bias refers to the bias which is displayed in the selection of items, rather than in the content of retrieved documents, namely content bias \citep{mowshowitz2002bias}. \citet{mowshowitz2002assessing,mowshowitz2005measuring} quantify instead only indexical bias by using precision and recall measures. Moreover, the researchers approximate the \emph{ideal} (i.e. norm) by the distribution produced by a collection of search engines to measure bias. Yet, this may not be a \emph{fair} bias evaluation procedure since the \emph{ideal} itself should be \emph{unbiased}, whereas the SERPs of search engines may actually contain \emph{bias}. Similarly, \citet{chen2006position} use the same method in order to quantify indexical and content bias, however, content analysis was performed by representing the SERPs with a weighted vector with different HTML tags without an in-depth analysis of the textual content. In this work, we evaluate content bias by analysing the textual contents of the Google and Bing SERPs, and we do not generate the \emph{ideal} relying on the SERPs of other search engines in order to measure bias in a more \emph{fair} way.
In addition to the categorisation of the content and indexical bias analysis, prior methods used in auditing algorithms to quantify bias can also be divided into three main categories as \emph{audience-based}, \emph{content-based}, and \emph{rater-based}.
\emph{Audience-based} measures focus on identifying the political perspectives of media outlets and web pages by utilising the interests, ideologies, or political affiliations of its users, e.g., likes and shares on Facebook~\citep{bakshy2015exposure}, based on the premise that readers follow the news sources that are closest to their ideological point of view~\citep{mullainathan2005market}.~\citet{lahoti2018joint} model the problem of ideological leaning of social media users and media sources in the liberal-conservative ideology space on Twitter as a constrained non-negative matrix-factorisation problem.
\emph{Content-based} measures exploit linguistic features in textual content; \cite{gentzkow2010drives} extract frequent phrases of the different political partisans (Democrats, Republicans) from the Congress Reports. Then, the researchers come with the metric of media slant index to measure US newspapers' political leaning. Finally, rater-based methods also exploit textual content and can be evaluated under the content-based methods. Unlike the content-based, the \emph{rater-based} methods use ratings of people for the sentiment, partisan or ideological leaning of content instead of analysing the textual content linguistically. Rater-based methods generally leverage crowd-sourcing to collect the labels for the content analysis. For instance, \citet{budak2016fair} quantify bias (partisanship) in US news outlets (newspapers and 2 political blogs) for 15 selected queries related to a wide range of controversial issues about which Democrats and Republicans argue. The researchers use MTurk as a crowd-sourcing platform to obtain the topic and political slant labels, i.e. being positive towards Democrats or Republicans, of the articles. Similarly, \citet{robertson2017auditing} use crowd-sourcing to score individual search results and \cite{diakopoulos2018vote} make use of the MTurk platform, i.e. rater-based approach, to get labels for the Google SERP websites by focusing on the content and apply an audience-based approach through utilising the prior work of~\cite{bakshy2015exposure} specifically for quantifying partisan bias. Our work follows a rater-based approach by making use of the MTurk platform for crowd-sourcing to analyse web search bias through stances and ideological leanings of the news articles instead of partisan bias in the textual contents of the SERPs.

There have been endeavors to audit partisan bias on web search. %provide a beneficial tool and have previously been examined to quantify partisan bias.
%analyse the effect of personalisation and location on search rankings~\cite{kliman2015locationbasedsearch, hannak2013personalizedsearch}.
%Apart from these studies, there have been recent works to audit partisan bias in non-personalised web search. 
\citet{diakopoulos2018vote} present four case studies on Google search results and to quantify partisan bias in the first page, they collect SERPs by issuing complete candidate names of the 2016 US presidential election as queries and utilise crowd-sourcing to obtain the sentiment scores of the SERPs. They found that Google presented a higher proportion of negative articles for Republican candidates than the Democratic ones. Similarly, \citet{robertson2017auditing} present a case study for the election and use a browser extension to collect Google and Yahoo search data for the election-related queries, then use crowd-sourcing to score the SERPs. The researchers also found a left-leaning bias and Google was more biased than Yahoo. In their follow-up work, they found a small but significant ranking bias in the standard SERPs but not due to personalisation~\citep{robertson2018auditing}. Similarly, researchers audit Google search after Donald Trump's Presidential inauguration with a dynamic set of political queries using auto-complete suggestions~\citep{robertson2018political}.~\citet{hu2019auditing} conduct an algorithm audit and construct a specific lexicon of partisan cues for  measuring political partisanship of Google Search snippets relative to the corresponding web pages. They  define the corresponding difference as bias for this particular use case without making a robust search bias evaluation of SERPs from the user's perspective. In this work, we introduce novel fairness-aware IR measures which involve rank information to evaluate content bias. For this, we use crowd-sourcing to obtain labels of the~\emph{news} SERPs returned towards the queries related to a wide-range of controversial topics instead of only political ones. With our robust bias evaluation measures, our main aim is to audit ideological bias in web search rather than solely partisan bias.

Apart from partisan bias, recent studies have investigated different types of bias for various purposes.~\citet{chen2018investigating} investigate gender bias in the various resume search engines, which are platforms that help recruiters to search for suitable candidates and use statistical tests to examine two types of indirect discrimination: individual and group fairness.
Similarly in another research study, authors investigate gender stereotypes by analyzing the gender distribution in image search results retrieved by Bing in four different regions~\citep{otterbacher2017competent}. Researchers use the query of `person' and the queries related to 68 character traits such as `intelligent person', and the results show that photos of women are more often retrieved for `emotional' and similar traits, whereas `rational' and related traits are represented by photos of men. In a follow-up work, researchers conduct a controlled experiment via crowd-sourcing with participants from three different countries to detect bias in image search results~\citep{otterbacher2018investigating}. Demographic information along with measures of sexism are analysed together and the results confirm that sexist people are less likely to detect and report gender biases in the search results.

~\citet{raji2019actionable} examine the impact of publicly naming biased performance results of commercial AI products in face recognition for directly challenging companies to change their products.~\citet{geyik2019fairness} present a fairness-aware ranking framework to quantify bias with respect to protected attributes and improve the fairness for individuals without affecting the business metrics. 
The authors extended the metrics proposed by~\citet{yang2017measuring}, of which we specified the limitations in Section~\ref{subsec:fairness_evaluation}, and evaluated their procedure using simulations with application to LinkedIn Talent Search.~\citet{vincent2019measuring} measure the dependency of search engines on user-created content to respond to queries using Google search and Wikipedia articles. In another work, researchers propose a novel metric that involves users and their attention for auditing group fairness in ranked lists~\citep{sapiezynski2019quantifying}.~\citet{gao2019fair} propose a framework that effectively and efficiently estimate the solution space where fairness in IR is modelled as an optimisation problem with fairness constraint. Same researchers work on top-k diversity fairness ranking in terms of statistical parity and disparate impact fairness and propose entropy-based metrics to measure the topical diversity bias presented in SERPs of Google using clustering instead of a labelled dataset with group information~\citep{gao2020toward}. Unlike to their approach, our goal is to quantify search bias in SERPs rather than topical diversity. For this, we use a crowd-labelled dataset, thereby to evaluate bias from the user's perspective with stance and ideological leanings of the documents.

%In many works, fairness in IR is modelled as an optimisation problem with fairness constraint. For instance, 

%\textcolor{red}{TODO: 2018-2019 cite papers} \\

%
In this context, we focus on proposing a new search bias evaluation procedure in ranked lists to quantify bias in the~\emph{news} SERPs. With the proposed robust fairness-aware IR measures, we also compare the relative bias of 
%we also make a bias-wise comparison of 
the two search engines through incorporating relevance and ranking information into the procedure without tracking the source of bias as discussed in Section \ref{sec:introduction}. Our procedure can be used for the source of bias analysis as well which we leave as future work.
%Consequently, to propose a stronger claim about algorithmic bias, one needs to fulfill a two-phased analysis of i. evaluating bias on SERPs, and ii. tracking the source of bias. We propose  that may be used in both steps of the analysis and in the scope of this work, we implement the first part, which is a preliminary step for the second phase, i.e., investigating the source of bias, to be completed in the follow-up work.

\section{Search Engine Bias Evaluation Framework}
\label{sec:bias_evaluation_measures}
%Article Title: Measuring Search Engines' Political Bias in Controversial Topics via Stance Bias
%Journal Name: Information Retrieval Journal
%Authors: Gizem Gezici, Aldo Lipani, Yucel Saygin, and Emine Yilmaz
%Corresponding Author Information: Gizem Gezici, Sabanci University, gizemgezici@sabanciuniv.edu

In this section we describe our  search bias evaluation framework. Then, we present the measures of bias and the proposed protocol to identify search bias. %To begin with, a set of symbols, functions, and labels used throughout this paper can be found in Table~\ref{table:notation}.

\subsection{Preliminaries}
\label{sec:preliminaries}

Our first aim is to detect bias with respect to the distribution of stances expressed in the contents of the SERPs. % and the term \emph{sentiment polarity} reflects the emotion  in a given text.

%Follows a set of symbols, functions, and labels used in this paper:
%\vspace{1.5em}

Let $\set{S}$ be the set of search engines and $\set{Q}$ be the set of queries about controversial topics. 
When a query $q \in \set{Q}$ is issued to a search engine $s \in \set{S}$, the search engine $s$ returns a SERP $r$. 
We define the stance of the $i$-th retrieved document $r_i$ with respect to $q$ as $j(r_i)$. A stance can have the following values:
\emph{pro}, % ($+$), 
\emph{neutral}, % ($\cdot$), 
\emph{against}, % (-) or 
\emph{not-relevant}. % ($\times$).

A document stance with respect to a topic can be:
\begin{itemize}
\item\textbf{pro} ($\lps$) when the document is in favour of the controversial topic. The document describes more the pro aspects of the topic;
\item\textbf{neutral} ($\lns$) when the document does not support or help either side of the controversial topic. The document provides an impartial (fair) description about the pro and cons of the topic;
\item\textbf{against} ($\las$) when the document is against the controversial topic. The document describes more the cons aspects of the topic;
\item\textbf{not-relevant} ($\lnr$) when the document is not-relevant with respect to the controversial topic.
\end{itemize}

For our analyses, we deliberately use recent \emph{controversial} topics in US that are the real debatable ones rather than the topics being possibly exposed to false media balance, which occurs when the media present opposing viewpoints as being more equal than the evidence supports, e.g. Flat Earth debate~\citep{FalseBalance, FalseBalance2}. Our topic set contains abortion, illegal immigration, gay marriage, and similar \emph{controversial} topics which comprise opposing points of view since complicated concepts concerning the identity, religion, political or ideological leaning are the actual points where search engines are more likely to provide biased results~\citep{noble2018algorithms} and influence people dramatically.
%Because of the polarized viewpoints, controversial queries may contain sentiment-bearing keywords and punctuation or imply a specific sentiment pole. Hence, bias in the retrieved results could also be explained by the sentiment-bearing keywords present in the search query itself. 
%In order to make sure that our analysis is not affected by such sentiment-bearing queries, we measure bias using \emph{all} queries in the query set \(Q\), as well as using a restricted set of \emph{non-polarized} queries denoted by \(Q_{nonPol}\) as reported in the experimental results.

%pro means the news article's author, if asked about the topic, would answer  positively, e.g. being in favour of the topic.
%Neutral means that the news article author is only informative in nature and provides pro and cons of both perspectives, pro and against.
%Against means the news article's author, if asked about the topic, would answer negatively, e.g. being against of the topic.

Our second aim is to detect bias with respect to the distribution of ideological leanings expressed in the contents of the SERPs. We do this by associating each query $q \in \set{Q}$ belonging to a controversial topic to one \emph{current} ideological leaning. Then, combining the stances for each $r_i$ and the associated ideological leaning of $q$ we can measure the ideological bias of the content of a given SERP, e.g.~if a topic belongs to a specific ideology and a document retrieved for this topic has a pro stance, we consider this document to be biased towards this ideology.
We define the ideological leaning of $q$ as $j(q)$. An ideological leaning can have the following values:
\emph{conservative}, % (\bc), 
\emph{liberal}, % (\rc), 
\emph{both or neither}. % (\wc).

A topic ideological leaning can be:
\begin{itemize}
\item \textbf{conservative} ($\lcp\,$) when the topic is part of the conservative policies. The conservatives are in favour of the topic;
\item \textbf{liberal} ($\llp\,$) when the topic is part of the liberal policies. The liberals are in favour of the topic;
\item \textbf{both or neither} ($\lbp\,$) when both or neither policies are either in favour or against the topic.
\end{itemize}

\begin{table*}
\centering
\caption{Symbols, functions, and labels used throughout the paper}
\begin{tabular}{ l l }
\hline
\hline
{Symbols}\\
\hline
\hline
$\sS$   & set of search engines. \\
$s$     & a search engine $s \in \sS$. \\
$\sQ$   & set of queries. \\
$q$     & a query $q \in \sQ$. \\
%$d$     & a document. \\
$r$     & a ranked list of the given SERP (list of retrieved documents). \\%(list of documents).
$r_i$   & the document in $r$ retrieved at rank $i$.\\
$|r|$ & size of $r$ (number of documents in the ranked list). % = [r_1, r_2, \dots, r_{|r|}]$.
\\
$n$     & number of documents considered in $r$ %by $f$ 
(cut-off).\\
%$G$     & a set of groups $g_i$ ($g_1 \cup g_2 \cup \dots \cup g_n = G$). \\
%$g_i$   & a group $g_i \in G$. \\ \hline
%\hline
\hline
\hline
{Functions}\\
\hline
\hline
%$s(q)$                  & returns $r$ given a query ($q$). \\ 
$j(r_i)$                & returns the label associated to $r_i$.\\
%$j(q)$                  & returns the label associated to $q$.\\
$f(r)$                  & an evaluation measure for SERPs.\\
%$f_g(r)$                & like $f$ but for $g$.\\
%$f_{-}(r)$             & like $f$ but for the against label.\\
%$[\text{Condition}]$    & returns 1 if the condition is true and 0 otherwise.\\
\hline
\hline
{Labels}\\
\hline
\hline
%$g_1$  & protected group. \\
%$g_2$  & unprotected group. \\ \hline
$\lps$ & pro stance. \\
$\lns$ & neutral stance. \\
$\las$ & against stance. \\ 
$\lnr$ & not-relevant stance. \\ \hline
$\lcp$ & conservative ideological leaning.
\\
%$\lnp$ & neutral political perspective. \\
$\llp$ & liberal ideological leaning. \\
$\lbp$ & both or neither ideological leanings. \\
%$+$ & label  .\\
%$-$ & against label.\\
\hline
\hline
\end{tabular}
\label{table:notation}
\end{table*}

For reference, Table~\ref{table:notation} shows a summary of all the symbols, functions and labels used in this paper. 

\subsection{Measures of Bias}
\label{sec:measuresbias}

Based on the aforementioned definition provided in Section \ref{sec:introduction}, bias can be quantified by measuring the degree of deviation of the distribution of documents from the \textit{ideal} one. To give a broad definition of an ideal list poses problems; but in the scope of this work for \emph{controversial} topics, we can mention the existence of bias in a ranked list retrieved by a search engine if the presented information \emph{significantly deviates} from true likelihoods~\citep{white2013beliefs}. 
As justified in Section~\ref{sec:introduction}, in the scope of this work we focus on~\emph{equality of output}, thus we accept the true likelihoods of different views as \emph{equal} rather than computing them from the corresponding base rates.
Therefore using the proposed definition reversely, we can assume that the \textit{ideal} list is the one that minimises the difference between two opposing views, which we indicate here as $\lps$ and $\las$ in the context of stances.
%form the \emph{unbiased} lists by \emph{randomly} generating the ranked document lists and accept them as \textit{IDEAL} lists for our framework. 
%Then for quantifying bias, we measure the difference between the ranked lists returned from the two search engines 
%, Bing and Google and the \textit{IDEAL} list in terms of the proposed sentiment-based metrics. The steps of the overall procedure for quantifying bias in SERPs is as follows. 
%
%Our aim is to detect the existence of bias in SERPs of a search engine as well as to compare the SERPs of two search engines in terms of bias with respect to controversial topics.
Formally, we measure the \emph{stance} bias in a SERP $r$ as follows:
\begin{linenomath*}
\begin{equation}\label{eq:bias}
    \beta_f(r) = f_{\lps}(r) - f_{\las}(r),
\end{equation}
\end{linenomath*}
where $f$ is a function that measures the likelihood of $r$ in satisfying the information need of the user about the view $\lps$ and the view $\las$. We note that \emph{ideological} bias is measured in the same way by transforming the stances of the documents into ideological leanings which will be explained in Section \ref{sec:crowdsourcing}.
Before defining $f$, from Eq.~\eqref{eq:bias}, we define the mean bias (MB) of a search engine $s$ as:
\begin{linenomath*}
\begin{equation*}\label{eq:expected_biasMB}
    \text{MB}_f(s, \sQ) = 
    \frac{1}{|\sQ|}
    \sum_{q \in \sQ} \beta_f(s(q)).
\end{equation*}
\end{linenomath*}
An unbiased search engine would produce a mean bias of $0$. 
A limitation of MB is that if a search engine is biased towards the $\lps$ view on one topic and bias towards the $\las$ view on another topic, these two contributions will cancel each other out. 
In order to avoid this limitation we also define the mean absolute bias (MAB), which consists in taking the absolute value of the bias for each $r$. Formally, this is defined as follows:
\begin{linenomath*}
\begin{equation}\label{eq:expected_biasMAB}
    \text{MAB}_f(s, \sQ) = 
    \frac{1}{|\sQ|}
    \sum_{q \in \sQ} |\beta_f(s(q))|.
\end{equation}
\end{linenomath*}
An unbiased search engine produces a mean absolute bias of $0$. Although this measure defined in Eq.~\eqref{eq:expected_biasMAB} solves the limitation of MB, MAB says nothing about towards which view the search engine is biased, making these two measures of bias complementary.

In IR the likelihood of $r$ in satisfying the information need of users is measured via retrieval evaluation measures. Among these measures we selected 3 \emph{utility-based} evaluation measures. This class of evaluation measures quantify $r$ in terms of its worth to the user and are normally computed as a sum of the information gain summed over the relevant documents retrieved by $r$. 
The 3 IR evaluation measures used in the following experiments are: 
P$@n$, RBP, and DCG$@n$.

P$@n$ for the $\lps$ view is formalised as in Eq. \eqref{eq:pc}. %the ratio between the relevant retrieved documents over the cut-off $n$. To quantify the bias towards one view we reformulate P$@n$ as the ratio between the relevant retrieved documents for the view over the cut-off $n$. For example, for the $+$ view, this is formalized as follows:
%
%\begin{equation}\label{eq:pc}
%    \text{P}_+@n = 
%    \frac{1}{n}\sum_{i=1}^n 
%    [j(r_i) = +]
%\end{equation}
%
%where the notation $[Condition]$ (aka Iverson bracket) returns 1 when the condition is verified and 0 when otherwise. $j(r_i)$ returns the stance of the document with respect to the topic $q$. 
However, differently from the previous definition of $j(r_i)$ where the only possible outcomes are $g_1$ and $g_2$ for the document $r_i$, here $j$ can return any of the label associated to a stance ($\lps$, $\lns$, $\las$, and $\lnr$). Hence, only pro and against documents, that are relevant to the topic, are taken into account, since $j(r_i)$ returns \emph{neutral} and \emph{not-relevant} when otherwise.
Substituting Eq.~\eqref{eq:pc} to Eq.~\eqref{eq:bias} we obtain the first measure of bias:
\begin{linenomath*}
\begin{equation*}\label{eq:bias_pc}
    \beta_{\text{P}@n}(r) = 
    \frac{1}{n}\sum_{i=1}^n 
    \left([j(r_i) = \lps] - [j(r_i) = \las]\right).
\end{equation*}
\end{linenomath*}
The main limitation of this measure of bias is that it has a weak concept of ranking, 
i.e.~the first $n$ documents contribute equally to the bias score. The next two evaluation measures overcome this issue by defining discount functions.

RBP weights every document based on the coefficients of a normalised geometric series with value $p \in ]0,1[$, where $p$ is a parameter of RBP. Similarly to what is done for P$@n$, we reformulate RBP to measure bias as follows:
\begin{linenomath*}
\begin{equation}\label{eq:rbpc}
    \text{RBP}_{\lps} = 
    (1-p) 
    \sum_{i=1} p^{i-1}[j(r_i) = \lps].
\end{equation}
\end{linenomath*}
Substituting Eq.~\eqref{eq:rbpc} to Eq.~\eqref{eq:bias} we obtain:
\begin{linenomath*}
\begin{equation*}\label{eq:bias_rbpc}
    \beta_{\text{RBP}}(r) = (1-p) \sum_{i=1} p^{i-1} \left(
    [j(r_i) = \lps] - [j(r_i) = \las]
    \right).
\end{equation*}
\end{linenomath*}
DCG$@n$, instead, weights each document based on a logarithmic discount function. Similarly to what is done for P$@n$ and RBP, we reformulate DCG$@n$ to measure bias as follows:
\begin{linenomath*}
\begin{equation}\label{eq:dcgc}
    \text{DCG}_{\lps}@n = \sum_{i=1}^{n} 
    \frac{1}{\log(i + 1)}[j(r_i) = \lps].
\end{equation}
\end{linenomath*}
Substituting Eq.~\eqref{eq:dcgc} to Eq.~\eqref{eq:bias} we obtain:
\begin{linenomath*}
\begin{equation}\label{eq:bias_dcg}
    \beta_{\text{DCG}@n}(r) = \sum_{i=1}^{n}
    \frac{1}{\log(i + 1)}\left(
    [j(r_i) = \lps] - [j(r_i) = \las]
    \right)
\end{equation}
\end{linenomath*}

Since we are evaluating web-users, for P$@n$ and DCG$@n$ we set $n=10$ and for RBP we set $p = 0.8$. This last formulation (Eq.~\eqref{eq:bias_dcg}), although it looks similar to the rND measure, it does not suffer from the four limitations introduced in Section \ref{subsec:fairness_evaluation}. 
In particular all these presented measures of bias:
1) do not focus on one group;
2) use a binary score associated to the document stance or ideological leaning, similar to the way these measures are used in IR when considering relevance; also like in IR  
3) can be computed at each rank; 
4) exclude non-relevant documents from the measurement of bias and; the framework
5) provides various user models associated to the 3 IR evaluation measures: P$@n$, DCG$@n$, and RBP.

\subsection{Quantifying Bias}
\label{sec:protocol}
Using the measures of bias defined in the previous section we quantify the bias of the two search engines, Bing and Google using the \emph{news versions} of these search engines. % against an \textit{ideal} stance distribution.
Then, we compare them thereof. Following, we describe each step of the proposed procedure used to quantify bias in SERPs.
\begin{itemize}
\item \textbf{News Articles in SERPs.}
We obtained the controversial queries
issued for searching from ProCong.org [2018] and applied
some filtering steps on the initial query set. After filtering,
the final query set size became 57. We submitted each query
in the final query set to the US News search engines of
Google and Bing using a US proxy. Then, we extracted the
whole corpus returned by both engines in response to all
the queries in the set. Note that the data collection process
was done in a controlled environment such that the queries
are sent to the search engines at the same time. For more
details about the selection of the queries and crawling the
SERPs, please refer to the previous phase of our analysis.
After having crawled all the SERPs returned from both engines and extracted their contents, we annotated the top 10
documents. We obtained the stance label of each document
with respect to the queries via crowd-sourcing. To label the
ideological leaning of queries, we also used crowd-sourcing.
To obtain the ideologies of documents, we transformed the
stance labels into ideologies based on the ideological leaning
of their corresponding queries. The details about our crowdsourcing campaigns as well as the transformation process
can also be found in the first phase of our analysis.

\item \textbf{Bias Evaluation.}
We compute the bias measures for every SERP with all three IR-based measures of bias: P$@n$, RBP, and DCG$@n$. 
We then aggregate the results using the two measures of bias, MB and MAB. % polarity scores from TextBlob in document and sentence-level as \(Pol_{S_j} ({q_i})\), and \(SentPol_{S_j}({q_i})\) respectively for all the queries \(q_i\). Then, we apply \(TRSFM\) function on these scores.
\item \textbf{Statistical Analysis.}
To identify whether the bias measured is not a byproduct of randomness, 
%for the bias measure MAB, 
we compute a one-sample t-test: 
the null hypothesis is that no difference exists and that the true mean is equal to zero. 
If this hypothesis is rejected, hence there is a significant difference and we claim that the evaluated search engine is biased.
%
%Then, to find out towards which view the search engine is biased and show this is not due to noise, we compute a one-sample t-test on the bias measure MB: the null hypothesis is that no difference exists (between two opposing views) and that the true mean is equal to zero.
%
%Additionally, we compare the difference in bias measured across the two search engines, for both of the measures of MB and MAB, using a two-tailed paired t-test: the null hypothesis is that the difference between the two true means is equal to zero.
%
%If this hypothesis is rejected, hence there is a significant difference and we claim that there is a difference in bias between the two search engines.
%
Then, we compare the difference in bias measured across the two search engines using a two-tailed paired t-test: the null hypothesis is that the difference between the two true means is equal to zero. If this hypothesis is rejected, hence there is a significant difference, we claim that there is a difference in bias between the two search engines.
\end{itemize}

\

\section{Experimental Setup}
\label{sec:experiments}
%Article Title: Measuring Search Engines' Political Bias in Controversial Topics via Stance Bias
%Journal Name: Information Retrieval Journal
%Authors: Gizem Gezici, Aldo Lipani, Yucel Saygin, and Emine Yilmaz
%Corresponding Author Information: Gizem Gezici, Sabanci University, gizemgezici@sabanciuniv.edu

\begin{table*}[!t]
\centering
%\vspace{2em}
\caption{All controversial topics, topics marked with red dots are conservative and blue for liberal}\label{table:topics}
%\vspace{1em}
\resizebox{\columnwidth}{!}{%
%\begin{adjustbox}{max width=\textwidth}
\begin{tabular}{cp{5cm}|cp{5cm}|cp{5cm}}
\hline
\hline
\hspace{2pt}\bc\hspace{-4pt}&\textbf{Abortion}: Should Abortion Be Legal? &
\hspace{2pt}\bc\hspace{-4pt}&\textbf{Alternative Energy vs. Fossil Fuels}: Can Alternative Energy Effectively Replace Fossil Fuels? &
\hspace{2pt}\bc&\hspace{-4pt}\textbf{Animal Testing}: Should Animals Be Used for Scientific or Commercial Testing? \\\hline
\rc\hspace{-4pt}&\textbf{Banned Books}: Should Parents or Other Adults Be Able to Ban Books from Schools and Libraries? &
\hspace{2pt}\bc\hspace{-4pt}&\textbf{Bill Clinton}: Was Bill Clinton a Good President? &
\hspace{2pt}\bc\hspace{-4pt}&\textbf{Born Gay? Origins of Sexual Orientation}: Is Sexual Orientation Determined at Birth? \\\hline
\hspace{2pt}\wc\hspace{-4pt}&\textbf{Cell Phones Radiation}: Is Cell Phone Radiation Safe? &
\hspace{2pt}\bc\hspace{-4pt}&\textbf{Climate Change}: Is Human Activity Primarily Responsible for Global Climate Change? &
\hspace{2pt}\wc\hspace{-4pt}&\textbf{College Education Worth It?}: Is a College Education Worth It? \\\hline
\hspace{2pt}\rc\hspace{-4pt}&\textbf{Concealed Handguns}: Should Adults Have the Right to Carry a Concealed Handgun? & 
\hspace{2pt}\rc\hspace{-4pt}& \textbf{Corporal Punishment}: Should Corporal Punishment Be Used in K-12 Schools? &
\hspace{2pt}\rc\hspace{-4pt}&\textbf{Corporate Tax Rate \& Jobs}: Does Lowering the Federal Corporate Income Tax Rate Create Jobs? \\\hline
\hspace{2pt}\rc\hspace{-4pt}&\textbf{Cuba Embargo}: Should the United States Maintain Its Embargo against Cuba? &
\hspace{2pt}\wc\hspace{-4pt}&\textbf{Daylight Savings Time}: Should the United States Keep Daylight Saving Time? &
\hspace{2pt}\wc\hspace{-4pt}&\textbf{Drinking Age - Lower It?}: Should the Drinking Age Be Lowered from 21 to a Younger Age? \\\hline
\hspace{2pt}\rc\hspace{-4pt}&\textbf{Drone Strikes Overseas}: Should the United States Continue Its Use of Drone Strikes Abroad? &
\hspace{2pt}\wc\hspace{-4pt}&\textbf{Drug Use in Sports}: Should Performance Enhancing Drugs (Such as Steroids) Be Accepted in Sports? &
\hspace{2pt}\rc\hspace{-4pt}&\textbf{Electoral College}: Should the United States Use the Electoral College in Presidential Elections? \\\hline
\hspace{2pt}\bc\hspace{-4pt}&\textbf{Euthanasia \& Assisted Suicide}: Should Euthanasia or Physician-Assisted Suicide Be Legal? &
\hspace{2pt}\wc\hspace{-4pt}&\textbf{Vaping E-Cigarettes}: Is Vaping with E-Cigarettes Safe? &
\hspace{2pt}\bc\hspace{-4pt}&\textbf{Felon Voting}: Should Felons Who Have Completed Their Sentence (Incarceration, Probation, and Parole) Be Allowed to Vote? \\\hline
\hspace{2pt}\wc\hspace{-4pt}&\textbf{Fighting in Hockey}: Should Fighting Be Allowed in Hockey? &
\hspace{2pt}\bc\hspace{-4pt}&\textbf{Gay Marriage}: Should Gay Marriage Be Legal? &
\hspace{2pt}\wc\hspace{-4pt}&\textbf{Gold Standard}: Should the United States Return to a Gold Standard? \\\hline
\hspace{2pt}\wc\hspace{-4pt}&\textbf{Golf - Is It a Sport?}: Is Golf a Sport? &
\hspace{2pt}\bc\hspace{-4pt}&\textbf{Illegal Immigration}: Should the Government Allow Immigrants Who Are Here Illegally to Become US Citizens? &
\hspace{2pt}\bc\hspace{-4pt}&\textbf{Israeli-Palestinian Two-State Solution}: Is a Two-State Solution (Israel and Palestine) an Acceptable Solution to the Israeli-Palestinian Conflict?\\\hline
\hspace{2pt}\wc\hspace{-4pt}&\textbf{Lowering the Voting Age to 16}: Should the Voting Age Be Lowered to 16? &
\hspace{2pt}\bc\hspace{-4pt}&\textbf{Medical Marijuana}: Should Marijuana Be a Medical Option? &
\hspace{2pt}\wc\hspace{-4pt}&\textbf{Milk - Is It Healthy?}: Is Drinking Milk Healthy for Humans? \\\hline
\hspace{2pt}\bc\hspace{-4pt}&\textbf{Minimum Wage}: Should the Federal Minimum Wage Be Increased? &
\hspace{2pt}\bc\hspace{-4pt}&\textbf{National Anthem Protest}: Is Refusing to Stand for the National Anthem an Appropriate Form of Protest? &
\hspace{2pt}\bc\hspace{-4pt}&\textbf{Net Neutrality}: Should Net Neutrality Be Restored? \\\hline
\hspace{2pt}\bc\hspace{-4pt}&\textbf{Obamacare}: Obamacare
Is the Patient Protection and Affordable Care Act (Obamacare) Good for America? &
\hspace{2pt}\bc\hspace{-4pt}&\textbf{Obesity a Disease?}: Is Obesity a Disease? &
\hspace{2pt}\wc\hspace{-4pt}&\textbf{Olympics}: Are the Olympic Games an Overall Benefit for Their Host Countries and Cities? \\\hline
\hspace{2pt}\wc\hspace{-4pt}&\textbf{Penny - Keep It?}: Should the Penny Stay in Circulation? &
\hspace{2pt}\wc\hspace{-4pt}&\textbf{Police Body Cameras}: Should Police Officers Wear Body Cameras? &
\hspace{2pt}\rc\hspace{-4pt}&\textbf{Prescription Drug Ads}: Should Prescription Drugs Be Advertised Directly to Consumers? \\\hline
\hspace{2pt}\bc\hspace{-4pt}&\textbf{Prostitution - Legalize It?}: Should Prostitution Be Legal? &
\hspace{2pt}\bc\hspace{-4pt}&\textbf{Right to Health Care}: Should All Americans Have the Right (Be Entitled) to Health Care? &
\hspace{2pt}\rc\hspace{-4pt}&\textbf{Ronald Reagan}: Was Ronald Reagan a Good President? \\\hline
\hspace{2pt}\bc\hspace{-4pt}&\textbf{Sanctuary Cities}: Should Sanctuary Cities Receive Federal Funding? &
\hspace{2pt}\rc\hspace{-4pt}&\textbf{School Uniforms}: Should Students Have to Wear School Uniforms? &
\hspace{2pt}\bc\hspace{-4pt}&\textbf{School Vouchers}: Are School Vouchers a Good Idea? \\\hline
\hspace{2pt}\wc\hspace{-4pt}&\textbf{Social Media}: Are Social Networking Sites Good for Our Society? &
\hspace{2pt}\rc\hspace{-4pt}&\textbf{Social Security Privatization}: Should Social Security Be Privatized? & 
\hspace{2pt}\rc\hspace{-4pt}&\textbf{Standardized Tests}: Is the Use of Standardized Tests Improving Education in America? \\\hline
\hspace{2pt}\bc\hspace{-4pt}&\textbf{Student Loan Debt}: Should Student Loan Debt Be Easier to Discharge in Bankruptcy? &
\hspace{2pt}\wc\hspace{-4pt}&\textbf{Tablets vs. Textbooks}: Should Tablets Replace Textbooks in K-12 Schools? & 
\hspace{2pt}\bc\hspace{-4pt}&\textbf{Teacher Tenure}: Should Teachers Get Tenure? \\\hline
\hspace{2pt}\rc\hspace{-4pt}&\textbf{Under God in the Pledge}: Should the Words "Under God" Be in the US Pledge of Allegiance? &
\hspace{2pt}\bc\hspace{-4pt}&\textbf{Universal Basic Income}: Is Universal Basic Income a Good Idea? & 
\hspace{2pt}\wc\hspace{-4pt}&\textbf{Vaccines for Kids}: Should Any Vaccines Be Required for Children? \\\hline
\hspace{2pt}\bc\hspace{-4pt}&\textbf{Vegetarianism}: Should People Become Vegetarian? &
\hspace{2pt}\wc\hspace{-4pt}&\textbf{Video Games and Violence}: Do Violent Video Games Contribute to Youth Violence? & 
\hspace{2pt}\wc\hspace{-4pt}&\textbf{Voting Machines}: Do Electronic Voting Machines Improve the Voting Process? \\
\hline\hline
\end{tabular}
%\end{adjustbox}
}
\end{table*}

In this section we provide a description of our experimental setup based on the proposed method as defined in Section \ref{sec:protocol}. 
%\textcolor{gray}{The software used to run these experiments is provided at the following web-link: \emph{anonymised}.}
\subsection{Material}
\label{sec:material}

We obtained all the controversial topics from ProCon.org~\citeyearpar{ProCon}. ProCon.org is a non-profit charitable organisation that provides an online resource for search on controversial topics. ProCon.org selects the topics that are controversial and important to many US citizens by also taking the readers' suggestions into account. 
We collected all 74 controversial topics with their topic questions from the website. Then, we applied three filters on these topics for practical reasons without deliberately selecting any topics.
The first filter selects only the \textit{polar} questions, also known as yes-no questions because they have no different sides for the analysis.
This filter decreased the topic set size from 74 to 70. 
The second filter removes the topics that do not contain up-to-date information in their topic pages provided by ProCon.org since they are not \emph{recent} controversial topics and would not return up-to-date results. With the second filter, the number of topics became 64. 
Lastly, the third filter only includes the topics if both search engines return results for the corresponding topic questions, otherwise the comparison analysis would not be possible. After the last filter, the final topic set became the size of 57. 
Table~\ref{table:topics} contains the full list of controversial topic titles with questions used in this study.

We used the topic questions of these 57 topics for crawling. For example, the topic question of the topic title `abortion' is `Should Abortion Be Legal?'.
The topic questions reflect the main debate on the corresponding controversial topics and we used them as they are (i.e. including upper-cased characters, without removing punctuation, etc.) for querying the search engines. %on the 31th of July, 2018
%and {\color{red}used them as they are (i.e. including upper-cased characters, without removing punctuation etc.). We didn't use the topics but topic questions??}

%Website to update the flowchart: https://www.draw.io
%Open the template in the website named as "Copy of CIKM_FlowChart_Recent.drawio" in the figures folder.

 %We had two separate crowdsourcing tasks to detect political perspective of a given topic and stance of a given document. 

We collected the news search results in \emph{incognito mode} %without including 
to avoid any personalisation effect. Thus, the retrieved SERPs are not specific to anyone, but (presumably) general to US users. We submitted each topic question to US News search engines of Google and Bing using a US proxy. 
Since we used the \emph{news versions} of the two search engines, sponsoring results which may affect our analysis did not appear in the news search results at all.
Then, we firstly crawled the URLs of the retrieved results for the same topic question to minimise the time lags between the search engines since %documents retrieved by them for the same
the SERP of the same topic may vary over time. Subsequently, we extracted the textual contents of the top-10 documents using the crawled URLs. By this way, the time span between the SERPs of Google and Bing for each controversial topic (whole corpus) became 2-3 minutes on average. Moreover, before starting the crawling process, we firstly made some experiments with a small set of topics (different from the topic set provided in the paper) in the news search as well as default search and did not observe significant changes especially in the top-10 documents of the news search even in 10-15 minutes time lags. This indicates that the news search is less dynamic than default search and we believe that the 2-3 minutes of time lags would not drastically affect the search results.

\begin{figure*}[!t]
  \centering  
  \includegraphics[width=0.9\textwidth]{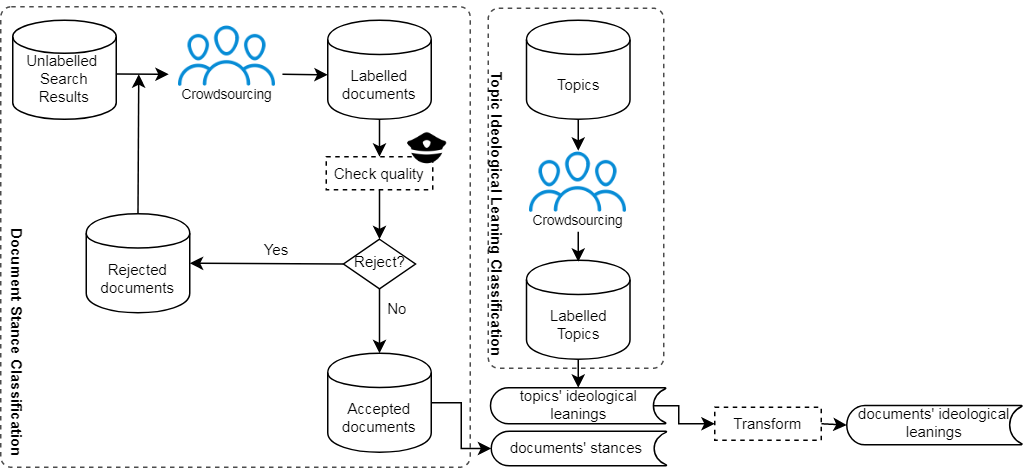}
  \caption{Flow-chart of the crowd-sourcing campaigns}
  \label{fig:flowchart}
\end{figure*}

\subsection{Crowd-sourcing Campaigns}
\label{sec:crowdsourcing}
%\textcolor{red}{Add details of the protocol}

The end-to-end process of obtaining stances and ideological leanings is shown in the flow-chart in Figure~\ref{fig:flowchart}. The emphasised (dotted) parts of the flow-chart show the steps of the Document Stance Classification (DSC) and Topic Ideological Leaning Classification (TILC).

\begin{comment}
\begin{table}[!htb]
    \begin{minipage}{.5\linewidth}
    \centering
    \caption{Crowd-workers Agreement}
    \begin{tabular}{cccc}
        \hline\hline
        Campaign & Inter-rater & Fleiss-Kappa \\
        \hline
        Document Stance & 0.4968 & 0.3500  \\
        Topic Political Perspective & 0.5281 & 0.3478  \\
        \hline\hline
    \end{tabular}
    \label{tab:agreement}
    \end{minipage}%
    \begin{minipage}{.5\linewidth}
    \centering
    % Engine 1 is better (Bing)
    \caption{Performance of the search engines. P-values of a two-tailed paired t-test computed between engine 1 and 2}
    \begin{tabular}{cccc}
        \hline\hline
        & P@10 & RBP & DCG@10 \\
        \hline
        Engine 1 & 0.8509 & 0.7708 & 3.9114 \\
        Engine 2 & 0.7404 & 0.6886 & 3.4773 \\
        \hline
        p-value & $< 0.001$ & $< 0.001$ & $< 0.01$ \\
        \hline\hline
    \end{tabular}
    \label{tab:performance}
    \end{minipage} 
\end{table}
\end{comment}

The DSC process inputs unlabelled top-10 search results, crawled by the data collection procedure described in Section~\ref{sec:material}, and outputs the stance labels of all these documents via crowd-sourcing with respect to the topic questions ($\sQ$) used to retrieve them. 
As displayed in the flow-chart, the TILC process uses crowd-sourcing to output the ideological leanings of all topic questions ($\sQ$). Then, the accepted stance labels of all documents, acquired from the DSC process are transformed into ideological leaning labels based on the assigned ideology of their corresponding topic questions. The steps of obtaining document labels in stance and ideological leaning detection are described below.

To label the stance of each document with respect to the topic questions ($\sQ$) we used crowd-sourcing. We selected MTurk as a crowd-sourcing platform. In this platform, to obtain high quality crowd-labels task properties were set as follows. Since the topics are mostly related to US, we selected crowd-workers only from US. Moreover, we tried to find qualified and experienced workers by setting the following thresholds: Human Intelligence Task (HIT) approval rate percentage should be greater than 95\% and number of HITs approved should be greater than 1000 for each worker. We set the wage as 0.15\$ and time allowed was 30 minutes per HIT. Each document was judged by three crowd-workers. 

%left-lower-right-upper
\begin{figure}[!t]
    \centering
    {\includegraphics[width=0.5\textwidth,trim={0.1cm 0.2cm 2cm 1.3cm},clip]{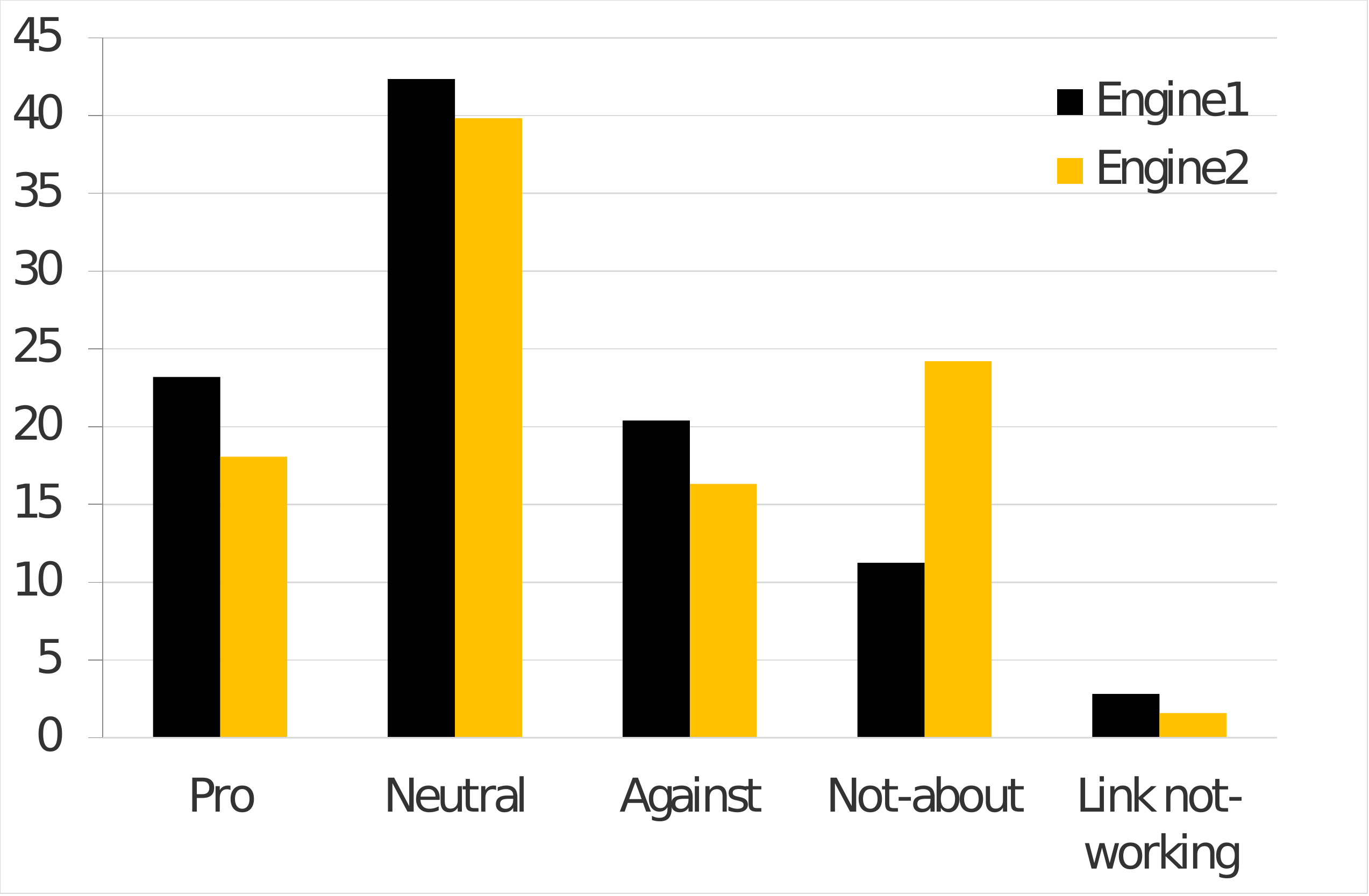}}
    \caption{Percentages of the document stance labels annotated by crowd-workers}
    \label{fig:stancelabel}
\end{figure}

To classify the stance of a document we asked crowd-workers to label, given a controversial topic question, the stance of a document in pro, neutral, against, not-relevant, or link not-working. Before the task was assigned, instructions were given to a worker in three groups from general to specific. Initially, workers were provided an overview of the stance detection task, then steps of the task were listed, i.e. read the topic question, open the news article link etc., and finally, rules and tips were displayed. This last part contained definitions of having a pro, neutral or against stance as given in Section~\ref{sec:preliminaries} above. Additionally, we included a clue for workers saying that title of the article may give you a general idea about the stance, however it is not sufficient to determine its overall viewpoint and then request workers to read also the rest of the article. Apart from these, at the end of the page we put a warning and informed the workers that some of the answers were known to us and we may reject their HITs, i.e. single, self-contained task for a worker, based on evaluation. Then, in the following page a HIT was shown to the worker with a topic question (query), link to the news article whose stance will be determined by  /reminding the main question of the stance detection task.

In order to obtain reliable annotations, we first annotated a randomly chosen set of documents later used to check the quality of crowd-labels as specified in the warning to the workers. With these expert labels, we rejected low quality annotations and requested new labels for those documents. This iterative process continued until we obtained all the document labels. %, in other words, the final corpus contained all the documents originally retrieved. 
At the end of this iterative process, for the sake of label reliability, we computed two agreement scores on the approved labels for document stance detection reported in Table~\ref{tab:agreement}. The  reported inter-rater agreement scores are the percent agreements between the corresponding annotators. We looked at pairwise agreement; put 1 if there is an agreement and 0, otherwise. Then we computed the mean for the fractions. Reported Kappa score for document stance classification is considered \emph{fair} agreement. Previously, researchers reported a Kappa score of the inter-rater agreement between experts (0.385) instead of crowd-workers for the same task, i.e. document stance classification in SERPs towards a different query set which includes controversial topics as well as popular products, by claiming MTurk workers had difficulty with the task~\citep{alam2014analyzing}. Although our task seems to be more challenging, i.e. the queries are only about controversial issues, our reported Kappa score for MTurk workers is comparable to their expert agreement score, which we believe to be sufficient due to the subjective nature and difficulty of the task.

The distribution of the accepted stance labels for the search results of each search engine is displayed in Figure~\ref{fig:stancelabel}. One may argue that for a query about a controversial topic issued to a news search engine, its SERP would mostly contain controversial articles that support one dominant viewpoint towards a given topic. Hence, informational pages or articles adequately discussing different viewpoints of the topic, i.e. documents that have a neutral stance, would never get a chance to be included in the analysis. However, the distribution in Figure~\ref{fig:stancelabel} refutes this argument by showing that the majority of the labels for both search engines is actually \textit{neutral}.

To identify the ideological leaning of each topic, we again used crowd-sourcing as displayed in Figure~\ref{fig:flowchart}. We asked the crowd-workers to classify each topic as: conservative, liberal, or both or neither. To get high quality annotations also for topic ideology detection, worker properties were set as the same with the stance detection. We again selected crowd-workers only from US. The wage per HIT was set as 0.1\$ and the time allowed was 5 minutes. Similarly to the stance detection, in the informational page we gave an overview, listed the steps and lastly provided the rules \& tips. For this task, last part contained the ideological leaning definitions as given in Section~\ref{sec:preliminaries}. Additionally, we requested the workers to evaluate the ideological leaning of a given topic based on the current ideological climate and warned them related to the rejection of their HITs as before. In the next page, the workers were shown a HIT with a topic question (query), i.e. one of the main debates of the corresponding topic, and asked the worker the following: \emph{Which ideological group would answer favourably to this question?}. The topics assigned to conservative or liberal leanings have been decided based on the judgment of five annotators with majority-voting. The leanings of the topics are shown in Table~\ref{table:topics}. Two agreement scores computed on the judgments for ideological leaning detection are also reported in Table~\ref{tab:agreement}.

\begin{table}[!t]
    %\vspace{1em}
    \centering
    \caption{Crowd-workers Agreement}
    \begin{tabular}{cccc}
        \hline\hline
        Campaign & Inter-rater & Fleiss-Kappa \\
        \hline
        Document Stance & 0.4968 & 0.3500  \\
        Topic Ideological Leaning & 0.5281 & 0.3478  \\
        \hline\hline
    \end{tabular}
    \label{tab:agreement}
\end{table}

\begin{table}[!t]
    %\vspace{1em}
    \centering
    % Engine 1 is better (Bing)
    \caption{Performance of the search engines, p-values of a two-tailed \\ paired t-test computed between engine 1 and 2}
    \begin{tabular}{cccc}
        \hline\hline
        & P@10 & RBP & DCG@10 \\
        \hline
        Engine 1 & 0.8509 & 0.7708 & 3.9114 \\
        Engine 2 & 0.7404 & 0.6886 & 3.4773 \\
        \hline
        p-value & $< 0.001$ & $< 0.001$ & $< 0.01$ \\
        \hline\hline
    \end{tabular}
    \label{tab:performance}
    %\vspace{1em}
\end{table}

To map the stance from the \emph{pro-to-against} to the \emph{conservative-to-liberal}, we applied a simple transformation to the documents. This transformation is needed because there may be documents which have a pro stance, for example, towards \emph{abortion} and \emph{Cuba embargo}. Though these documents have the same stance, they have different ideological leanings since having a pro stance on \emph{abortion} implies a \textit{liberal leaning}, whereas a pro stance on \emph{Cuba embargo} implies a \emph{conservative leaning}. 
For some topics (as in the case of Cuba embargo), we can directly interpret the \emph{pro-to-against} stance labels of search results as \emph{conservative-to-liberal} ideological leaning labels while for other topics (as in the case of the abortion) as \emph{liberal-to-conservative}. 
On the other hand, for those topics such as \emph{vaccines for kids}, which crowded label resulted in both or neither, 
the conservative-to-liberal or liberal-to-conservative transformation was not meaningful and therefore eliminated by our analysis. We note that within budget constraints, the crowd-sourcing protocol was designed to obtain crowd-labels with high-quality by labelling (expert) the random sample of documents, applying iterative process and majority voting on these labels.

\subsection{Results}
%Article Title: Measuring Search Engines' Political Bias in Controversial Topics via Stance Bias
%Journal Name: Information Retrieval Journal
%Authors: Gizem Gezici, Aldo Lipani, Yucel Saygin, and Emine Yilmaz
%Corresponding Author Information: Gizem Gezici, Sabanci University, gizemgezici@sabanciuniv.edu

In Table \ref{tab:performance} we present the performance of the two search engines. This is measured over all the topics. A document is considered relevant when classified as pro, against, or neutral. The difference for all evaluation measures is statistically significant.

\begin{table*}[!t]
    \centering
    \caption{
    Stance bias of the search engines, p-values of a two-tailed \\ paired t-test computed between engine 1 and 2 
    %P-values of two-tailed paired t-test computer between engine 1 and 2.
    %$\ast$ indicates non statistical significance %while $\dag$ (p-value < 0.001) statistical significance 
    %of a one-sample t-test.
    }
    \begin{tabular}{cccccccc}
        \hline\hline
        & & P@10 & & RBP & & DCG@10 & \\
        \hline
        \multirow{3}{*}{MB} 
        & Engine 1 & 0.0281 && 0.0197 && 0.1069 &\\	
        & Engine 2 & 0.0175	&& 0.0271 && 0.1142 &\\
        \cline{2-8}
        & p-value & $>0.05$ & & $>0.05$ & & $>0.05$ & \\
        \hline
        \multirow{3}{*}{MAB} 
        & Engine 1 & 0.2596	&& 0.2738 &&	1.3380 &\\
        & Engine 2 & 0.2246 &&	0.2266 &&	1.0789 &\\
        \cline{2-8}
        & p-value & $>0.05$ & & $>0.05$ & & $>0.05$ & \\
        \hline\hline
    \end{tabular}
    \label{tab:stance}
\end{table*}

\begin{table*}[!t]
    \centering
    \caption{Ideological bias of the search engines, p-values of a two-tailed \\ paired t-test computed between engine 1 and 2 %P-values of two-tailed paired t-test. 
    %$\ddag$ (p-value < 0.005) and $\dag$ (p-value < 0.05) signify statistical significance on a one-sample t-test.
    }
    \hspace{-0.7em}\begin{tabular}{cccccccc}
        \hline\hline
        & & P@10 & & RBP & & DCG@10 & \\
        \hline
        \multirow{3}{*}{MB} 
        & Engine 1 & -0.1368 & & -0.1247 & & -0.6290 & \\	
        & Engine 2 & -0.1289 & & -0.1386 & & -0.6591 &\\
        \cline{2-8}
        & p-value & $>0.05$ & & $>0.05$ & & $>0.05$ &\\
        \hline
        \multirow{3}{*}{MAB} 
        & Engine 1 & 0.2579 && 0.2894 && 1.3989 & \\
        & Engine 2 & 0.2184	&& 0.2158 && 1.0456 &  \\
        \cline{2-8}
        & p-value & $>0.05$ & & $<0.05$ & & $<0.05$ \\
        \hline\hline
    \end{tabular}
    \label{tab:political}
    %\vspace{1.2em}
\end{table*}

In Table \ref{tab:stance} we present the stance bias of the search engines.
Note that for all the three measures of bias, P@10, RBP and DCG@10, lower value is better which means lower bias in the scope of this work as opposed to their corresponding classic IR measures.
All MB and MAB scores are positive for all three IR evaluation measures. %The one-sample t-test computed on MBs are non-statistically significant and denoted as (*). 
Also, the differences between the two search engines for both MB and MAB measures are statistically not significant and it is shown with the two-tailed pair t-test on these measures.
In Table \ref{tab:political} we show the ideological bias. Similarly to Table \ref{tab:stance}, lower is better since we use the same measures of bias.
This table is similar to Table \ref{tab:stance}. Unlike the Table \ref{tab:stance}, all MB scores are negative while all MAB scores are positive for all three IR evaluation measures. %The one-sample t-test computed on MBs are statistically significant for all the measures with different confidence values. 
The two-tailed paired t-test computed on MBs to compare the difference in bias between engine 1 and engine 2, this is statistically not significant. Nonetheless, the two-tailed test on MABs is statistically not significant for the measure P@10; but it is statistically significant for the measures RBP and DCG@10.

%for the measure of P@10 is non-statistically significant, whereas for the measures of RBP and DCG@10, it is statistically significant and denoted as ($\dag$). Also here when comparing the bias between engine 1 and engine 2 this is not statistically significant. However, when measuring the bias between engine 1 and engine 2 for the measures RBP and DCG@10, it is statistically significant. Moreover, the political bias measured by MAB is statistically significant on a one-sample t-test.

%left-lower-right-upper
\begin{figure}[!t]
\vspace{-1.2em}
\centering
%\hspace{-1.8em}
    \begin{minipage}{0.45\textwidth}
        \centering
        \vspace{0.6em}
        \includegraphics[width=\textwidth,trim={2.5cm 6.5cm 4.5cm 6.5cm},clip]{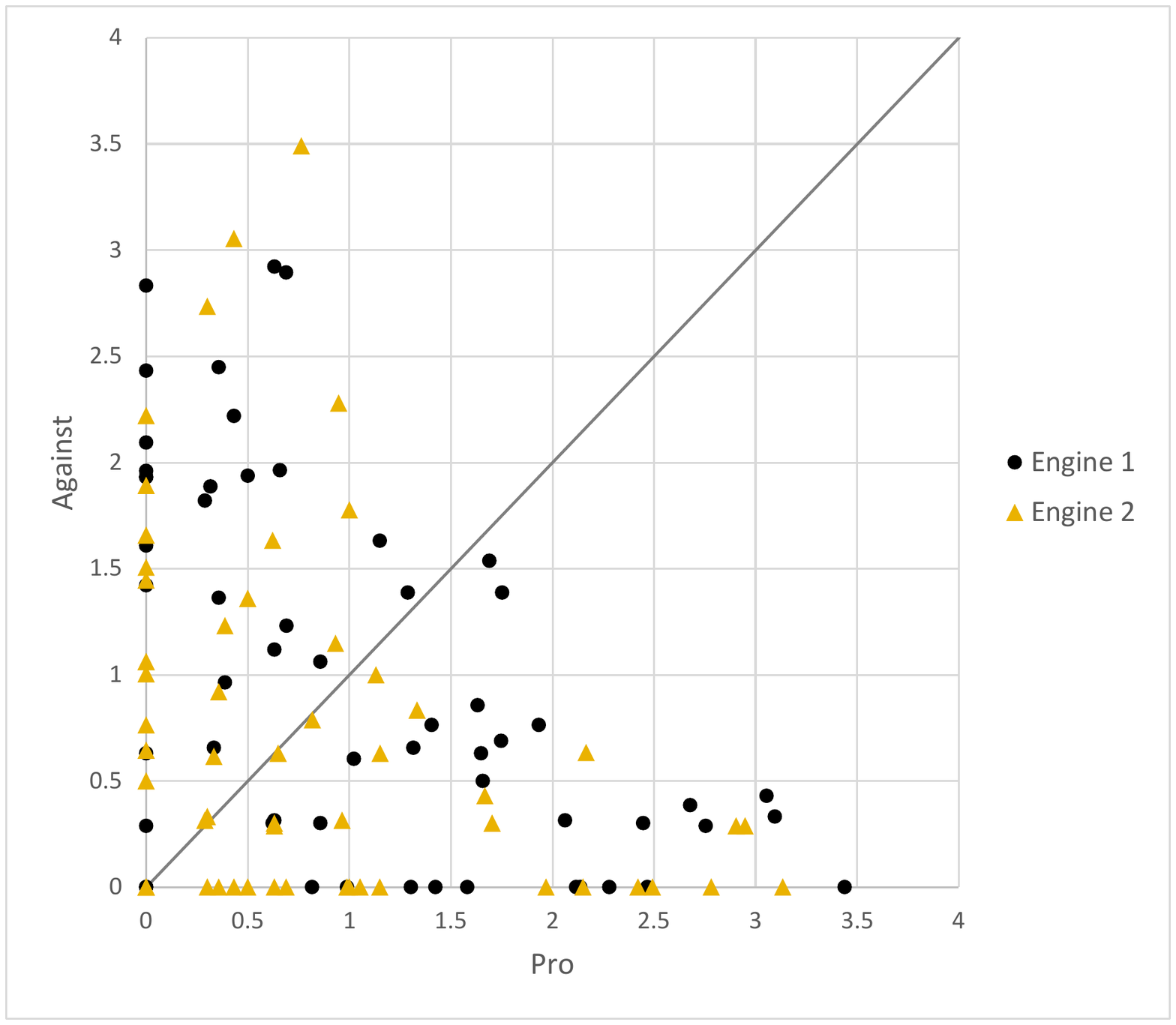}
        \caption{$DCG_{\lps}@10$ against $DCG_{\las}@10$ measured on stances -- black points for engine 1 and yellow points for engine 2}
        \label{fig:dcgstance}
\end{minipage}
\hspace{1.8em}
    \begin{minipage}{0.45\textwidth}
        \centering
        \vspace{1.4em}
        %\hspace{-1.8em}
        \includegraphics[width=\textwidth,trim={2.5cm 6.5cm 4.5cm 6.5cm},clip]{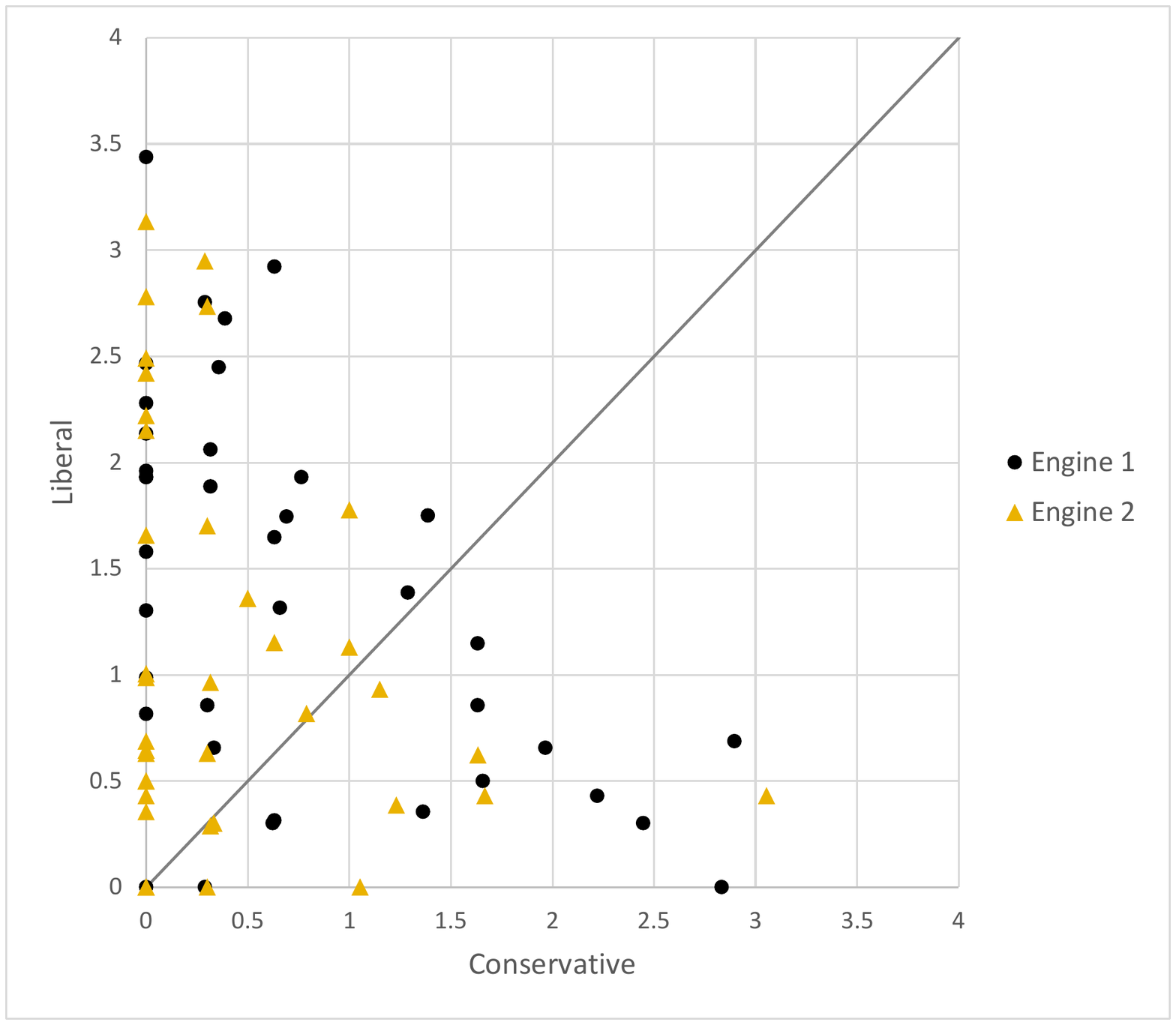}
        \caption{$DCG_{\lcp}@10$ against $DCG_{\llp}@10$ measured on ideological leanings -- black points for engine 1 and yellow points for engine 2}
    \label{fig:dcgpol}
\end{minipage}
\vspace{2em}
\end{figure}

\begin{figure}[!t]
\centering
%\hspace{-1.8em}
\begin{minipage}{.43\textwidth}
  \centering
    %\hspace{-1em}
    \includegraphics[width=\textwidth,trim={2.5cm 5.25cm 2cm 5cm},clip]{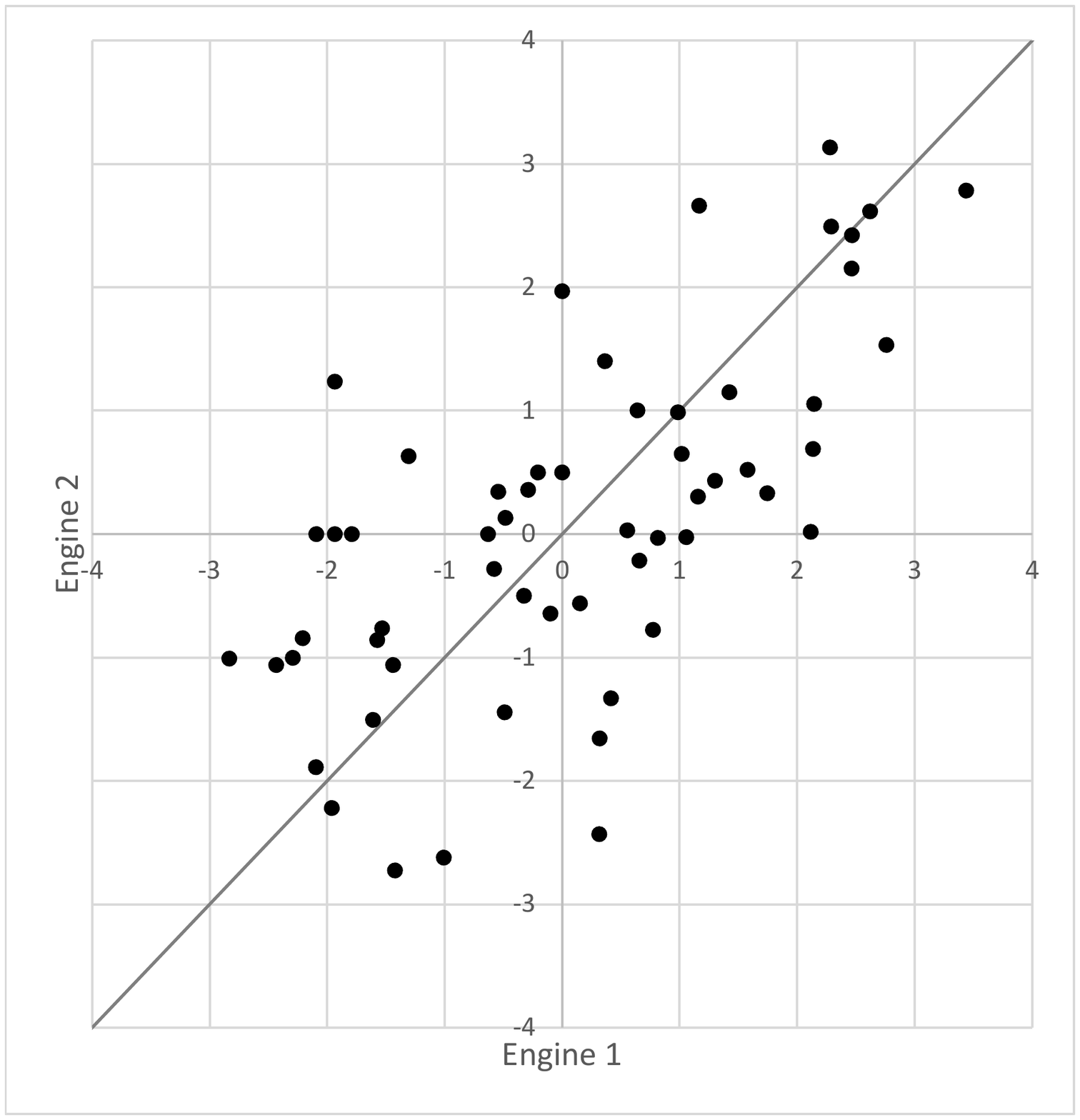}
    \caption{$\beta_{DCG@10}$ measured on stances, where positive is $\lps$ and negative is $\las$}
    \label{fig:dcgdiffstance}
\end{minipage}
\hspace{2.8em}
\begin{minipage}{.43\textwidth}
    \centering
    \vspace{0.1em}
    %\hspace{-1em}
    {\includegraphics[width=1\textwidth,trim={2.5cm 5.25cm 2cm 5cm},clip]{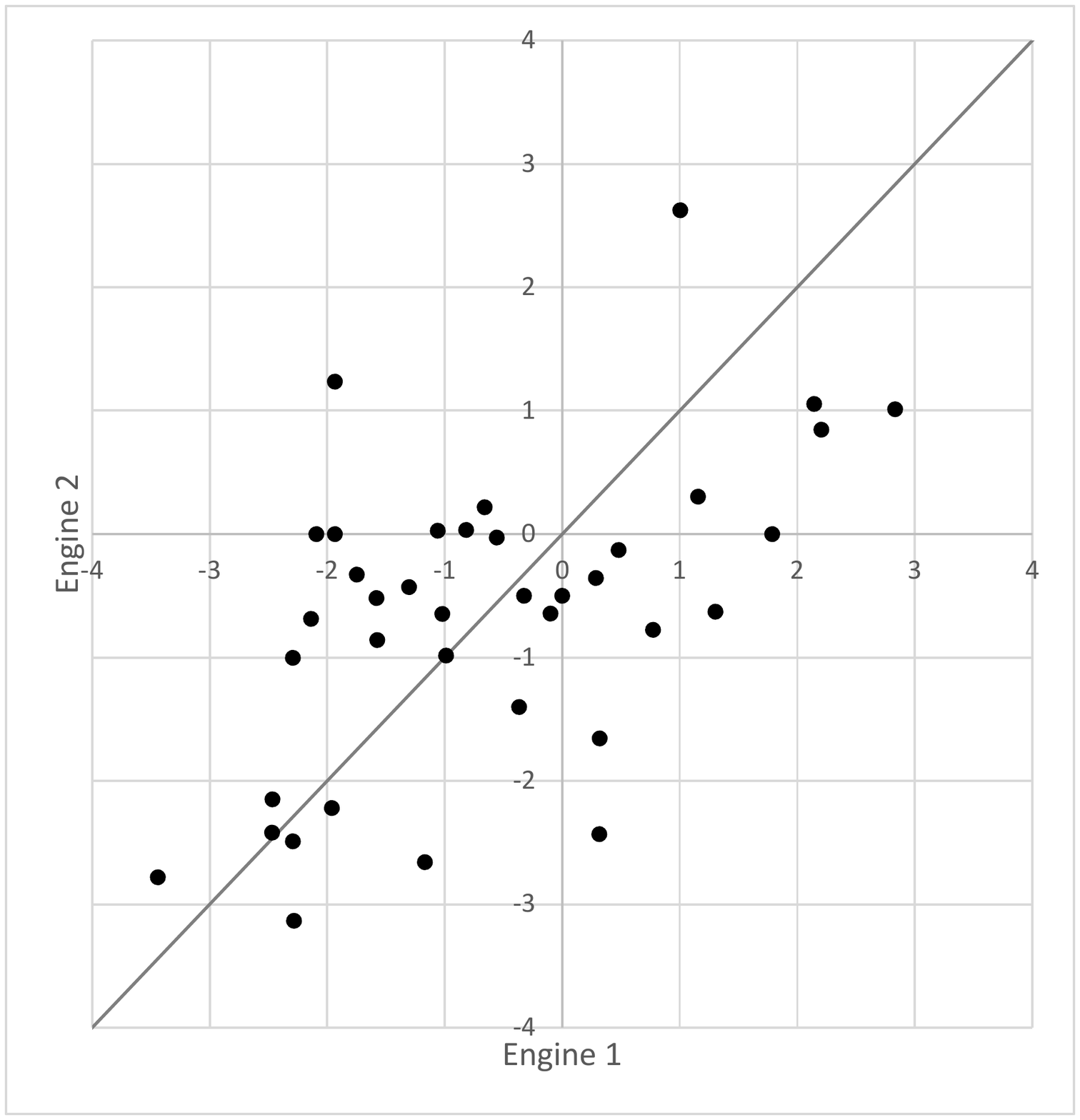}}
    \caption{$\beta_{DCG@10}$ measured on leanings, where positive is $\lcp$ and negative is $\llp$}
    \label{fig:dcgdiffpol}
\end{minipage}
%\vspace{-1.5em}
\end{figure}

In Figure \ref{fig:dcgstance} we show how the topic-wise SERPs distribute over the pro-against stance space for the measure DCG@10. The x-axis is the pro stance score ($DCG_{\lps}@10$) and the y-axis is the against stance score ($DCG_{\las}@10$). 
Each point corresponds to the overall SERP score of a topic. Black points are those SERPs retrieved by engine 1 and yellow points are those retrieved by engine 2.

In Figure \ref{fig:dcgdiffstance} we compare the overall stance bias score ($\beta_{DCG@10}$), i.e. difference between the pro and against stance scores, of SERPs for each topic measured on the two search engines. The x-axis is engine 1 and the y-axis is engine 2. The points in positive coordinates denote the topics whose SERPs are overall biased towards the pro stance, negative coordinates are for the against stance. 

Figure \ref{fig:dcgpol} and Figure \ref{fig:dcgdiffpol} are similar to Figure \ref{fig:dcgstance} and Figure \ref{fig:dcgdiffstance} but instead of measuring the stance bias we measure the ideological bias in the former case. Therefore, Figure \ref{fig:dcgpol} displays how the overall SERPs of topics distribute over the conservative-liberal ideological space for the measure DCG@10. Similarly, in Figure \ref{fig:dcgdiffpol} we compare the overall ideological bias score ($\beta_{DCG@10}$), i.e. difference between the conservative and liberal leaning scores, of the SERPs where the points in positive coordinates stand for the topics that are biased towards the conservative leaning, negative coordinates are for the liberal.

%\begin{figure}[!ht]
%\centering
%\captionsetup{justification=centering}
%    \includesvg[width=0.53\textwidth]{figures/AvgPol}
%    \captionsetup{font=scriptsize}
%   \caption[short]{The distribution of \textit{all} \\ queries on \(AVG_{Senti}@10\)}
%    \label{fig:avg}
%\end{figure}

%\begin{figure}[h!]
%    \begin{minipage}[t]{0.52\textwidth}
%    \includesvg[width=\textwidth]{figures/NDCG.svg}
%    \captionsetup{font=scriptsize}
%    \caption[short]{The distribution of \textit{all} \\ queries on \(NDCG_{Senti}@10\)}
%    \label{fig:ndcg}
%    \end{minipage}
%    \hfill
%    \begin{minipage}[t]{0.5\textwidth}
%    \includesvg[width=\textwidth]{figures/AP.svg}
%    \captionsetup{font=scriptsize}
%    \caption[short]{The distribution of \textit{all} \\ queries on \(AP_{Senti}@10\)}
%    \label{fig:ap}
%    \end{minipage}
%\end{figure}

%{\color{red} Something about Q1}

\section{Discussion}
\label{sec:discussion}
%Article Title: Measuring Search Engines' Political Bias in Controversial Topics via Stance Bias
%Journal Name: Information Retrieval Journal
%Authors: Gizem Gezici, Aldo Lipani, Yucel Saygin, and Emine Yilmaz
%Corresponding Author Information: Gizem Gezici, Sabanci University, gizemgezici@sabanciuniv.edu

Before investigating the existence of bias in SERPs, we initially compared the retrieval performances of two search engines. In Table~\ref{tab:performance} we observe that the performance of the two search engines is high but engine 1 is better than engine 2 -- their difference is statistically significant. This is verified across all three IR evaluation measures. %This suggest that the search engines  %All three evaluation measures support that Engine 1 performs better in retrieving relevant documents, RBP with 99\% confidence (i.e., $\alpha = 0.01$) and other two measures even with higher confidence values.

%Symbols in the table were obtained from: http://www.apsstylemanual.org/apsHouseStyle/handEditing/tables.htm
%Finally, any statistical information indicated in the table by symbols should be described. APS has a specific order for the symbols; they are described in the order of  * † ‡ §. If the symbols are out of order or if other symbols have been used within the table, then you will have to reorder or alter them. If fewer than 4 symbols have been used, then change to letters (in alphabetical order).

%%%%% RQ1 and RQ2

Next, we verify if the search engines return biased results in terms of document stances (RQ1) and if so, we further investigate if the engines suffer from the same level of bias (RQ2) that the difference between the engines are not statistically significant. 
%
%For this, we computed MB and MAB scores for all three evaluation measures. 
%
In Table~\ref{tab:stance} %we observe that
all MB scores are positive and regarding the RQ1, the engines seem to be biased towards the pro stance. 
%the engines seem to be biased towards the pro stance (all MB scores are positive - RQ1). This means that both search engines are biased towards the same stance (RQ2).
%
We applied the one-sample t-test on MB scores to check the existence of stance bias, i.e. if the true mean is different from zero, as mentioned in Section~\ref{sec:protocol}. However, these biases are statistically not significant which means that this expectation may be the result of noise -- there is not a systematic stance bias, i.e. preference of one stance with respect to the other. 
Based on MAB scores, we can observe that both engines suffer from an absolute bias. However, the difference between the two engines is shown to be non-significant with the two-tailed t-test. %This bias is significant for all evaluation measures. 
%shows the p-values for both experiments all three evaluation measures of MB and MAB give consistent results. 
%
These results show that both search engines are not biased towards a \emph{specific} stance in returning results since there is no statistically significant difference from the \emph{ideal} distribution. Nonetheless, for both engines there exists an absolute bias which can be interpreted as the expected bias for a topic question. These empirical findings imply that the search engines are biased for some topics towards the pro stance and for others towards the against stance.
%Yet, all the three measures of MB (denoted as $\ast$).

%
The results are displayed in Figure~\ref{fig:dcgstance}. This figure refers to the values used to compute the MAB score of the DCG$@10$ column. %scores distribution of each query referring to the Table~\ref{tab:stance} and 
It shows that the difference between the pro and against stances of both engines for topics is uniformly distributed. To note that, no topic can be located on the up-right area of the plot because the sum of their coordinates is bounded by the maximum possible DCG$@10$ score. Moreover we observe that topics are distributed similarly across the engines.
This is also confirmed by Figure~\ref{fig:dcgdiffstance} where we can observe that the stance bias scores ($\beta_{DCG@10}$), i.e. the differences between DCG$@10$ scores for the pro stance and DCG$@10$ scores for the against stance, of topics are somehow balanced between the up-right quadrant and the low-left quadrant. Moreover, these two quadrants are the area of agreement in stance between the two engines. The other two quadrants contain those topics where the engines disagree. Here we can conclude that the engines agree with each other in the majority of cases.

%engine 1 and engine 2 are denoted as points. 
%
%We also put a line of equality ($y=x$) (i.e. Bing and Google scores for 
%The specified measures for the axis labels are \emph{equal} if they appear on the line) for comparison. 
%If a point lies above the line, this means that the 
%Google 
%y-axis label score of the corresponding query is higher than x-axis, and vice-versa. 
%
%The up-right (I: +/+) and low-left (III: -/-) regions have some points (the engines agree on the pro (+) and against (-) stances), while the up-left (IV: -/+) and low-right (II: +/-) have fewer points (the engines disagreed on) that the search engines mostly agree with each other. 
%

%%%%% RQ3
Lastly, we investigate if the search engines are biased in the ideology space (RQ3).
Looking at MB scores in Table~\ref{tab:political} we observe that both search engines seem to be biased towards the same ideological leaning -- liberal (all MB scores are negative). Unlike the stance bias, one sample t-test on MB scores show that these expectations are statistically significant with different confidence values, i.e. p-value < 0.005 across all three IR measures for engine 2; whereas the same confidence value on P$@10$ for engine 1 and p-value < 0.05 on RBP and DCG$@10$. These results indicate that both search engines are biased towards the same leaning which is liberal.
%the p-values for two aggregating measures of bias in terms of political perspective.
Comparing the two search engines on MB scores, we observe that their differences are statistically not significant, which means that the observed difference may be the result of random noise.
%
%However, the results are significant and all evaluation measures agreed on the fact that both search engines are \textit{biased}. 
%
%However, they are \textit{biased} as much since the difference between their bias scores (MB) of all three IR-based measures are non-significant. 
%
Based on MAB, since all MAB scores are positive we can also observe that both engines suffer from an absolute bias. However, in contrast with what observed for the stance bias, this time there is a difference in expected ideological bias between the two search engines. For RBP and DCG@10 the difference between the engines is statistically significant. This finding and the different user models that these evaluation measures model suggest that the perceived bias by the users may change based on their behaviour. A user that always inspects the first 10 results (as modelled by P$@10$) may perceive the same ideological bias between engine 1 and engine 2, while a less systematic user, which just inspects the top results, may perceive that engine 1 is more biased than engine 2. Moreover, comparing this finding with the performance of the engines, we can observe that the better performing engine is more biased than the worse performing one.

%On the other hand, there is a significant difference between the SERPs of engine 1 and engine 2 in terms of absolute bias (MAB), if we include strong concept of ranking in our analysis (i.e., with DCG and RBP measures), that Engine 1 is more \textit{biased}. 

%where the difference is non-significant for {\color{red}delta bias} but the difference both search engines are \textit{biased} and they are \textit{biased} as much in terms of political leanings if we include rank into our analysis. Moreover, our results show that Engine 1 (Bing) is more \textit{biased} and are significant with 95\% confidence (i.e., $\alpha = 0.05$). all evaluation measures also agreed on the fact that the two search engines have similar expected stance towards the same controversial topic because there is no significant difference between the SERPs of Bing and Google.

%Hence, we argue that both search engines tend to show biased results with respect to the controversial topics, and they tend to be biased in the same way. 

%In order to further compare the SERPs of the two search engines in stance and political perspective detection, 
%bias present in the two search engines, que
%we plotted the DCG and Diff DCG scores
%three sentiment-based metric scores 
%of all controversial queries in $\set{Q}$ to visualize the distribution of the values for Bing and Google in Figure~\ref{fig:dcgstance}, 
%Figure~\ref{fig:dcgdiffstance}, 
%Figure~\ref{fig:dcgpol}, and 
%Figure~\ref{fig:dcgdiffpol}. 
%
%We represented each query as a point; each point is a topic. 
%

Comparing Figure~\ref{fig:dcgpol} with Figure~\ref{fig:dcgstance} we observe that in Figure \ref{fig:dcgpol} the points look less uniformly distributed than in Figure \ref{fig:dcgstance}. Topics are mostly on the liberal side. Moreover, engine 2 has fewer points on the conservative side than engine 1. %shows the stance analysis per topic with the qcuery points of engine 1 and engine 2 which compare the DCG scores of two different stances as \emph{pro} and \emph{against}. 
%
%If a point lies above the line, this means that the \emph{against}-DCG score is higher than \emph{pro}-DCG for the corresponding query
%
%he following plots can be interpreted in a similar way.
Comparing Figure~\ref{fig:dcgdiffpol} with Figure~\ref{fig:dcgdiffstance}, we observe that the engines in Figure~\ref{fig:dcgdiffpol} are more biased towards the liberal side with respect to what observed in Figure~\ref{fig:dcgdiffstance}. Also, we observe that the engines mostly agree -- most of the points are placed on the up-right and low-left quadrants.

In conclusion, we find important to point out that it is not in the scope of this work to find the source of bias. As discussed in the introduction, bias may be a result of the input data, which may contain biases, or the search algorithm, which contains sophisticated features and specifically chosen algorithms that, although designed to be effective in satisfying information needs, may produce systematic biases. 
Nonetheless, we look at the problem from the user perspective and no matter where the bias comes from; the results are biased as described. Our findings seem to be consistent with prior works~\citep{robertson2017auditing, diakopoulos2018vote} that there exists liberal (left-leaning) partisan bias in SERPs; even in unpersonalised search settings~\citep{robertson2018auditing}. %Moreover, Epstein recently published a testimony~\citep{epsteingoogle} for the US Congress and based on the extensive experiments he has fulfilled so far, he stated that Google censored conservative content which legitimises Mr. Trump's argument.

\section{Limitations}
\label{sec:limitations}

This work has potential limitations. As stated in the introduction, we focus on a particular kind of bias, known as \emph{statistical parity}, or more generally known as \emph{equality of outcome} instead of \emph{equality of opportunity} which uses query-specific base rates. In the context of the controversial topics where the document labels were obtained via crowd-sourcing, this bias measure, i.e. requiring equal representation of stances instead of query-specific base rates, made our experiments feasible. %Then, our choice of ideal ranking requires equal representation of stances instead of query-specific base rates which made our experiments feasible. 
This is firstly because, not all of the query questions in our list have certain answers based on scientific facts, i.e. some of them are subjective queries.
In investigating the equality of opportunity, queries can be further categorized as subjective and objective on top of our evaluation framework. For the objective queries, expert labels can be obtained and used as base rates, then search results can be evaluated by taking into account these base rates. Please note that our evaluation framework could better be applied to the controversial queries from the public's perspective mainly where the goal is to have balanced SERPs instead of skewed results. We believe that some queries should be handled with a different framework since those queries are not intrinsically controversial such as \emph{Is Holocaust real?} - there is only one correct answer without the need of a discussion.

Besides, the identification of the stance for the full ranking list is currently too expensive to get annotated via crowd-sourcing. To tackle this issue, a machine learning model can help us to automate the process of obtaining the stance labels. 
Another potential limitation is that some queries may not be real user queries. Nonetheless, we extracted the queries directly from their topic pages of the ProCon.org~\citeyearpar{ProCon} along with the topics. We deliberately did not change the queries to avoid any interference/bias from our side on the results. 
In this work, we did not make a domain-specific selection of the topics, or apply any filtering as subjective/objective, rather we accepted them as \emph{controversial} topics from the general public's perspective which is the main scope of this work.

Apart from these, crowd-workers' own personal biases may affect the labelling process. For this reason, we tried to mitigate these biases by i. asking the workers to annotate stances rather than ideologies to make their judgment more objective, and ii. aggregating the final judgment coming from multiple workers.
Additionally, our analysis refers to a specific point in time where the data was collected. To enable reproducibility and an easier comparison of these results at some point in the future, we made our dataset publicly available. %The bias analysis allows us to investigate bias in search results.
Lastly, we note that this bias analysis can only be used as an indicator of potentially biased ranking algorithms because it is not enough in order to track the source of bias. In the scope of this work, we did not investigate the source of bias that may come from the data (input bias) or from the ranking mechanism (algorithmic bias) of the corresponding search engines. 
Despite these potential limitations, we believe that our work is a good attempt to evaluate bias in search results with new bias measures and a dataset crawled specifically for the search bias evaluation. Since the bias analysis is very complex, we deliberately limited our scope and only focused on the bias analysis of \emph{recent} controversial topics in \emph{news} search. Nonetheless, all these limitations lead us to numerous interesting future directions.

\section{Conclusion \& Future Work}
\label{sec:conclusion}

In this work we introduced new bias evaluation measures and a generalisable evaluation framework %using these measures 
to address the issue of web search bias in news search results. 
We applied the proposed framework to measure stance and ideological bias in the SERPs of Bing and Google as well as compare their relative bias towards controversial topics.
%We tested our framework and reported the experimental evaluation on the SERPs of Bing and Google in response to controversial topics to detect possible bias and to compare SERPs with respect to two search biases.
Our initial results show that both search engines seem to be unbiased when considering the document stances and \emph{ideologically} biased when considering the document ideological leanings.
In this work, we intended to analyse SERPs without the effect of personalisation. Thus, these results highlight that search biases exist even though the personalization effect is minimized
%without personalisation 
and that search engines can empower users by being more accountable. %Thus, the results also indicate that even without personalisation, the returned results for a query may contain bias.
%Article Title: Measuring Search Engines' Political Bias in Controversial Topics via Stance Bias
%Journal Name: Information Retrieval Journal
%Authors: Gizem Gezici, Aldo Lipani, Yucel Saygin, and Emine Yilmaz
%Corresponding Author Information: Gizem Gezici, Sabanci University, gizemgezici@sabanciuniv.edu

%Even though Engine 1 may be slightly more biased than Engine 2, the difference does not seem to be considerably significant.
In the scope of this work we did not investigate the source of bias which we left as future work, therefore the results can be seen as a potential indicator.
% that may come from the data (input bias) or from the ranking mechanism (algorithmic bias) of the corresponding search engines 
%which we left as future work. 
In our experiments, we gathered document stances via crowd-sourcing. Thus, the obvious future work in this direction is to use automatic stance detection methods instead of crowd-sourcing to obtain the document labels, thereby evaluating bias in the whole corpus of retrieved SERPs to track the source of bias. 
Moreover, investigating the workers' bias in a follow-up work would be interesting since it is very difficult to remove all biases in practice.
In this work, we focus on \emph{equality of outcome}; but using another bias measure, \emph{equality of opportunity} which takes into account the corresponding group proportions, i.e. query-specific base rates, in the population would be an alternative follow-up work. We plan to categorize queries as subjective and objective, then modify the \emph{ideal} ranking definition specifically for the objective queries based on the corpus distributions.
The bias analysis for the objective queries, particularly the ones related to the critical domains such as health search, can be investigated further on top of our evaluation framework which we believe to be an interesting follow-up work.
Furthermore, we plan to study the effect of localization and personalization, i.e. how much the stances and ideological leanings varied across users or the echo chamber effect, on SERPs, then incorporate that study into our bias evaluation framework in the future.

%\newpage

\section*{Compliance with Ethical Standards}
Author Emine Yilmaz previously worked as a research consultant for Microsoft Research and she is currently a research consultant for Amazon Research.

\section*{Acknowledgements}
We thank the reviewers for their comments. This work has been funded by the EPSRC Fellowship titled "Task Based Information Retrieval", grant reference number EP/P024289/1 and the visiting researcher programme of The Alan Turing Institute.

\bibliographystyle{besjournals}
\bibliography{main}

\end{document}